  \documentclass[iop]{emulateapj}
  \usepackage{apjfonts}

\shorttitle{ASTEROSEISMIC AGE AND RADIUS OF KIC 11026764}
\shortauthors{METCALFE ET AL.}

\begin{document}

\title{A precise asteroseismic age and radius for the evolved Sun-like 
star KIC~11026764}

\author{
   T.~S.~Metcalfe\altaffilmark{1},
   M.~J.~P.~F.~G.~Monteiro\altaffilmark{2},
   M.~J.~Thompson\altaffilmark{3,1},
   J.~Molenda-\.Zakowicz\altaffilmark{4},
   T.~Appourchaux\altaffilmark{5},
   W.~J.~Chaplin\altaffilmark{6},
   G.~Do\u{g}an\altaffilmark{7},
   P.~Eggenberger\altaffilmark{8},
   T.~R.~Bedding\altaffilmark{9},
   H.~Bruntt\altaffilmark{10},
   O.~L.~Creevey\altaffilmark{11,12},
   P.-O.~Quirion\altaffilmark{13},
   D.~Stello\altaffilmark{9},
   A.~Bonanno\altaffilmark{14},
   V.~Silva~Aguirre\altaffilmark{15},
   S.~Basu\altaffilmark{16},
   L.~Esch\altaffilmark{16},
   N.~Gai\altaffilmark{16,17},
   M.~P.~Di~Mauro\altaffilmark{18},
   A.~G.~Kosovichev\altaffilmark{19},
   I.~N.~Kitiashvili\altaffilmark{20},
   J.~C.~Su\'arez\altaffilmark{21},
   A.~Moya\altaffilmark{22},
   L.~Piau\altaffilmark{23},
   R.~A.~Garc\'ia\altaffilmark{23},
   J.~P.~Marques\altaffilmark{24},
   A.~Frasca\altaffilmark{14},
   K.~Biazzo\altaffilmark{25},
   S.~G.~Sousa\altaffilmark{2},
   S.~Dreizler\altaffilmark{26},
   M.~Bazot\altaffilmark{2},
   C.~Karoff\altaffilmark{6},
   S.~Frandsen\altaffilmark{7},
   P.~A.~Wilson\altaffilmark{27,28},
   T.~M.~Brown\altaffilmark{29},
   J.~Christensen-Dalsgaard\altaffilmark{7},
   R.~L.~Gilliland\altaffilmark{30},
   H.~Kjeldsen\altaffilmark{7},
   T.~L.~Campante\altaffilmark{2,7},
   S.~T.~Fletcher\altaffilmark{31},
   R.~Handberg\altaffilmark{7},
   C.~R\'egulo\altaffilmark{11,12},
   D.~Salabert\altaffilmark{11,12},
   J.~Schou\altaffilmark{19},
   G.~A.~Verner\altaffilmark{32},
   J.~Ballot\altaffilmark{33},
   A.-M.~Broomhall\altaffilmark{6},
   Y.~Elsworth\altaffilmark{6},
   S.~Hekker\altaffilmark{6},
   D.~Huber\altaffilmark{9},
   S.~Mathur\altaffilmark{1},
   R.~New\altaffilmark{31},
   I.~W.~Roxburgh\altaffilmark{32,10},
   K.~H.~Sato\altaffilmark{23},
   T.~R.~White\altaffilmark{9},
   W.~J.~Borucki\altaffilmark{34},
   D.~G.~Koch\altaffilmark{34},
   J.~M.~Jenkins\altaffilmark{35}
}
\altaffiltext{1}{High Altitude Observatory, National Center for Atmospheric Research, Boulder, CO 80307, USA}

\altaffiltext{2}{Centro de Astrof\'\i sica and DFA-Faculdade de Ci\^encias, Universidade do Porto, Portugal}

\altaffiltext{3}{School of Mathematics and Statistics, University of Sheffield, Hounsfield Road, Sheffield S3 7RH, UK}

\altaffiltext{4}{Astronomical Institute, University of Wroc\l{}aw, ul. Kopernika 11, 51-622 Wroc\l{}aw, Poland}

\altaffiltext{5}{Institut d'Astrophysique Spatiale, Universit\'e Paris XI -- CNRS (UMR8617), Batiment 121, 91405 Orsay Cedex, France}

\altaffiltext{6}{School of Physics and Astronomy, University of Birmingham, Edgbaston, Birmingham, B15 2TT, UK}

\altaffiltext{7}{Department of Physics and Astronomy, Aarhus University, DK-8000 Aarhus C, Denmark}

\altaffiltext{8}{Geneva Observatory, University of Geneva, Maillettes 51, 1290, Sauverny, Switzerland}

\altaffiltext{9}{Sydney Institute for Astronomy (SIfA), School of Physics, University of Sydney, NSW 2006, Australia}

\altaffiltext{10}{Observatoire de Paris, 5 place Jules Janssen, 92190 Meudon Principal Cedex, France}

\altaffiltext{11}{Instituto de Astrof\'{\i}sica de Canarias, E-38200 La Laguna, Spain}

\altaffiltext{12}{Departamento de Astrof\'{\i}sica, Universidad de La Laguna, E-38206 La Laguna, Spain}

\altaffiltext{13}{Canadian Space Agency, 6767 Boulevard de l'A\'eroport, Saint-Hubert, QC, J3Y 8Y9, Canada}

\altaffiltext{14}{INAF -- Osservatorio Astrofisico di Catania, Via S.Sofia 78, 95123 Catania, Italy}

\altaffiltext{15}{Max Planck Institute for Astrophysics, Karl Schwarzschild Str. 1, Garching, D-85741, Germany}

\altaffiltext{16}{Department of Astronomy, Yale University, P.O. Box 208101, New Haven, CT 06520-8101, USA}

\altaffiltext{17}{Beijing Normal University, Beijing 100875, P.R. China}

\altaffiltext{18}{INAF-IASF Roma, Istituto di Astrofisica Spaziale e Fisica Cosmica, via del Fosso del Cavaliere 100, 00133 Roma, Italy}

\altaffiltext{19}{HEPL, Stanford University, Stanford, CA 94305-4085, USA}

\altaffiltext{20}{Center for Turbulence Research, Stanford University, 488 Escondido Mall, Stanford, CA 94305, USA}

\altaffiltext{21}{Instituto de Astrof\'{\i}sica de Andaluc\'{\i}a (CSIC), CP3004, Granada, Spain}

\altaffiltext{22}{Laboratorio de Astrof\'{\i}sica, CAB (CSIC-INTA), Villanueva de la Ca\~nada, Madrid, PO BOX 78, 28691, Spain}

\altaffiltext{23}{Laboratoire AIM, CEA/DSM-CNRS, Universit\'e Paris 7 Diderot, IRFU/SAp, Centre de Saclay, 91191, Gif-sur-Yvette, France}

\altaffiltext{24}{LESIA, CNRS UMR 8109, Observatoire de Paris, Universit\'e Paris 6, Universit\'e Paris 7, 92195 Meudon Cedex, France}

\altaffiltext{25}{Arcetri Astrophysical Observatory, Largo Enrico Fermi 5, 50125 Firenze, Italy}

\altaffiltext{26}{Georg-August Universit\"{a}t, Institut f\"{u}r Astrophysik, Friedrich-Hund-Platz 1, D-37077 G\"{o}ttingen}

\altaffiltext{27}{Nordic Optical Telescope, Apartado 474, E-38700 Santa Cruz de La Palma, Santa Cruz de Tenerife, Spain}

\altaffiltext{28}{Institute of Theoretical Astrophysics, University of Oslo, P.O. Box 1029, Blindern, N-0315 Oslo, Norway }

\altaffiltext{29}{Las Cumbres Observatory Global Telescope, Goleta, CA 93117, USA}

\altaffiltext{30}{Space Telescope Science Institute, Baltimore, MD 21218, USA}

\altaffiltext{31}{Materials Engineering Research Institute, Sheffield Hallam University, Sheffield, S1 1WB, UK}

\altaffiltext{32}{Astronomy Unit, Queen Mary, University of London, Mile End Road, London, E1 4NS, UK}

\altaffiltext{33}{Laboratoire d'Astrophysique de Toulouse-Tarbes, Universit\'e de Toulouse, CNRS, 14 av E. Belin, 31400 Toulouse, France}

\altaffiltext{34}{NASA Ames Research Center, MS 244-30, Moffett Field, CA 94035, USA}

\altaffiltext{35}{SETI Institute, NASA Ames Research Center, MS 244-30, Moffett Field, CA 94035, USA}

%\email{travis@ucar.edu}
  \journalinfo{ACCEPTED for publication in the Astrophysical Journal}

\begin{abstract}

The primary science goal of the {\it Kepler Mission} is to provide a 
census of exoplanets in the solar neighborhood, including the 
identification and characterization of habitable Earth-like planets. The 
asteroseismic capabilities of the mission are being used to determine 
precise radii and ages for the target stars from their solar-like 
oscillations. Chaplin et al.~(2010) published observations of three bright 
G-type stars, which were monitored during the first 33.5~d of science 
operations. One of these stars, the subgiant KIC~11026764, exhibits a 
characteristic pattern of oscillation frequencies suggesting that it has 
evolved significantly. We have derived asteroseismic estimates of the 
properties of KIC~11026764 from {\it Kepler} photometry combined with 
ground-based spectroscopic data. We present the results of detailed 
modeling for this star, employing a variety of independent codes and 
analyses that attempt to match the asteroseismic and spectroscopic 
constraints simultaneously. We determine both the radius and the age of 
KIC~11026764 with a precision near 1\%, and an accuracy near 2\% for the 
radius and 15\% for the age. Continued observations of this star promise 
to reveal additional oscillation frequencies that will further improve 
the determination of its fundamental properties.

\end{abstract}


%%%%%%%%%%%%%%%%%%%%%%%%%%%%%%%%%%%%%%%%%%%%%%%%%%%%%%%%%%%%%%%%%%%%%%%%%%%

\section{INTRODUCTION}

% FIGURE 1 ---------------------------------------------------------------
  \begin{figure*}[t]
  \centerline{\includegraphics[angle=0,width=6.5in]{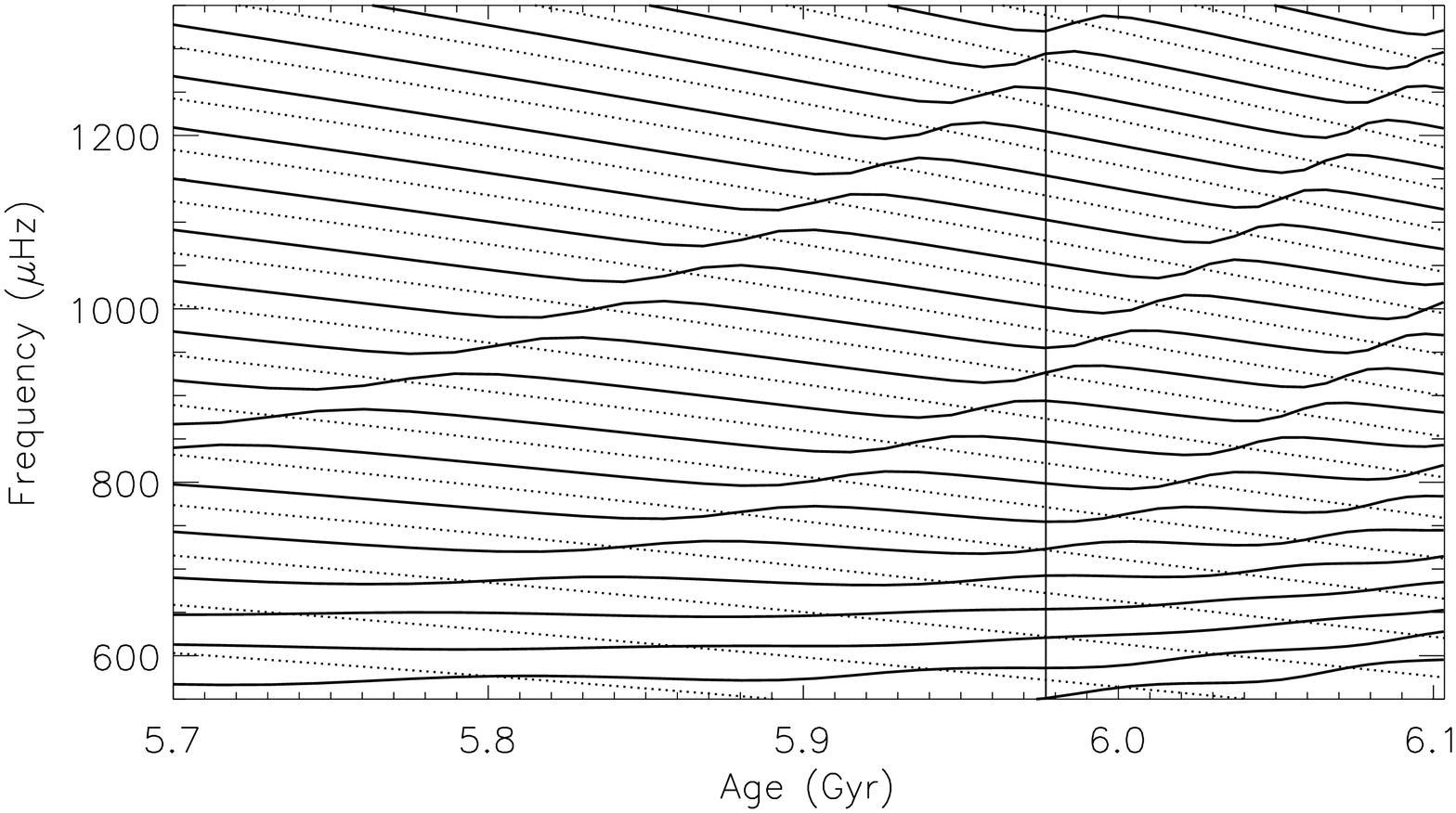}}
  \caption{Evolution of the $l=0$ (dotted) and $l=1$ (solid) 
  oscillation frequencies as a function of age for a representative stellar 
  model of KIC~11026764. The frequency separation between consecutive $l=1$ 
  modes during an avoided crossing is a strong function of the stellar age. 
  Note the prediction of a high-frequency avoided crossing above 
  1250~$\mu$Hz for Model AA, indicated by the vertical line.\label{fig1}\\}
  \end{figure*}
%-------------------------------------------------------------------------

In March 2009 NASA launched the {\it Kepler} satellite, a mission designed 
to discover habitable Earth-like planets around distant Sun-like stars. 
The satellite consists of a 0.95-m telescope with an array of digital 
cameras that will monitor the brightness of more than 150,000 solar-type 
stars with a precision of a few parts-per-million for 4-6 years 
\citep{bor07}. Some of these stars are expected to have planetary systems, 
and some of the planets will have orbits such that they periodically pass 
in front of the host star, causing a brief decrease in the amount of light 
recorded by the satellite. The depth of such a {\it transit} contains 
information about the size of the planet relative to the size of the host 
star.

Since we do not generally know the precise size of the host star, the 
mission design includes a revolving selection of 512 stars monitored with 
the higher cadence that is necessary to detect short period solar-like 
oscillations, allowing us to apply the techniques of asteroseismology 
\citep{jcd07,ack10}. Even a relatively crude analysis of such measurements 
can lead to reliable determinations of stellar radii to help characterize 
the extra-solar planetary systems discovered by the mission, and stellar 
ages to reveal how such systems evolve over time. For the asteroseismic 
targets that do not contain planetary companions, these data will allow a 
uniform determination of the physical properties of hundreds of solar-type 
stars, providing a new window on stellar structure and evolution.

Initial results from the Kepler Asteroseismic Investigation were presented 
by \cite{gil10a}, while a more detailed analysis of the solar-like 
oscillations detected in several early targets was published by 
\cite{cha10}. The latter paper includes observations of three bright 
(V$\sim$9) G-type stars, which were monitored during the first 33.5~d of 
science operations. One of these stars, the subgiant KIC~11026764 
($\equiv$ 2MASS J19212465+4830532 $\equiv$ BD+48~2882 $\equiv$ 
TYC~3547-12-1), exhibits a characteristic pattern of oscillation 
frequencies suggesting that it has evolved significantly.

In unevolved stars, the high radial order ($n$) acoustic oscillation modes 
(p-modes) with a given spherical degree ($l$) are almost evenly spaced in 
frequency. As the star evolves and the envelope expands, the p-mode 
frequencies gradually decrease. Meanwhile, as the star becomes more 
centrally condensed, the buoyancy-driven (g-mode) oscillations in the core 
shift to higher frequencies. This eventually leads to a range of 
frequencies where the nonradial ($l>0$) oscillation modes can take on a 
mixed character, behaving like g-modes in the core and p-modes in the 
envelope (``mixed modes''), with their frequencies shifted as they undergo 
so-called {\it avoided crossings} \citep{Osaki75,Aizenman77}. This 
behavior changes relatively quickly with stellar age, and propagates from 
one radial order to the next as the star continues to evolve (see 
Figure~\ref{fig1}). Consequently, those modes that deviate significantly 
from uniform frequency spacing yield a strong (though model-dependent) 
constraint on the age of the star \citep[e.g., see][]{jcd04}. Avoided 
crossings have been observed in the subgiant stars $\eta$~Boo 
\citep{kje95,kje03,car05} and $\beta$~Hyi \citep{bed07}, and possibly also 
in Procyon \citep{bed10} and HD~49385 \citep{deh10}. As noted by 
\cite{gil10a} and \cite{cha10} the dipole ($l=1$) modes observed in 
KIC~11026764 show the signature of an avoided crossing, raising the 
exciting possibility that detailed modeling of this star will provide a 
very precise determination of its age.

In this paper we derive the stellar age, radius and other characteristics 
of KIC~11026764 by matching both the observed oscillation frequencies from 
{\it Kepler} photometry and the best available spectroscopic constraints 
from ground-based observations. We describe the extraction and 
identification of the oscillation frequencies in \S\ref{SEC2}, and the 
analysis of ground-based data for spectroscopic constraints in 
\S\ref{SEC3}. In \S\ref{SEC4} we provide the details of the independent 
codes and analysis methods used for the fitting, and in \S\ref{SEC5} we 
describe our final modeling results. We summarize and discuss the broader 
significance of the results in \S\ref{SEC6}.

%%%%%%%%%%%%%%%%%%%%%%%%%%%%%%%%%%%%%%%%%%%%%%%%%%%%%%%%%%%%%%%%%%%%%%%%%%%

% TABLE 1 ----------------------------------------------------------------
  \tablewidth{0pt}
  \tabletypesize{\normalsize}
  \tablecaption{The minimal and maximal sets of observed oscillation frequencies for KIC~11026764.\label{tab1}}
  \begin{deluxetable*}{cccccccc}
  \tablehead{
  \colhead{}&\multicolumn{3}{c}{Minimal Frequency Set ($\mu$Hz)} &
  \colhead{}&\multicolumn{3}{c}{Maximal Frequency Set ($\mu$Hz)} \\
  \cline{2-4}\cline{6-8}
  \colhead{$n$\tablenotemark{*}}&\colhead{$l=0$}&\colhead{$l=1$}&\colhead{$l=2$}&
  \colhead{}&\colhead{$l=0$}&\colhead{$l=1$}&\colhead{$l=2$}
  }
  \startdata
  10 & $\cdots$ & $\cdots$ & $\cdots$ &
  &$\cdots$&$\cdots$&$615.49\pm0.45$\tablenotemark{b}\\
  11 & $ 620.79 \pm 0.29$\tablenotemark{c} & $ 653.80 \pm 0.23$ & $\cdots$           &
  &$620.42\pm0.37$\tablenotemark{c}&$654.16\pm0.39$&$670.23\pm1.21$\tablenotemark{a}\\
  12 & $\cdots$ & $\cdots$ & $\cdots$ &
  &$673.97\pm0.71$\tablenotemark{a}&$699.87\pm0.66$\tablenotemark{b}&$716.36\pm0.21$\tablenotemark{b}\\
  13 & $ 723.63 \pm 0.10$ & $ 755.28 \pm 0.26$ & $ 767.46 \pm 0.46$ &
  & $ 723.30 \pm 0.24$ & $ 754.85 \pm 0.23$ &  $ 769.16 \pm 0.71$  \\
  14 & $ 772.82 \pm 0.30$ & $ 799.96 \pm 0.40$ & $ 817.91 \pm 0.58$ &
  & $ 772.53 \pm 0.33$ & $ 799.72 \pm 0.23$ &  $ 818.81 \pm 0.30$  \\
  15 & $ 822.72 \pm 0.14$ & $ 847.57 \pm 0.30$ & $ 867.66 \pm 0.86$ &
  & $ 822.46 \pm 0.31$ & $ 846.88 \pm 0.34$ &  $ 868.31 \pm 0.24$  \\
  16 & $ 873.55 \pm 0.14$ & $ 893.48 \pm 0.33$ & $\cdots$ &
  &$873.30\pm0.27$&$893.52\pm0.20$&$919.31\pm0.52$\tablenotemark{a}\\
  17 & $ 924.53 \pm 0.37$ & $ 953.57 \pm 0.39$ & $ 969.77 \pm 0.36$ &
  & $ 924.10 \pm 0.29$ & $ 953.51 \pm 0.22$ &  $ 970.12 \pm 0.68$  \\
  18 & $ 974.59 \pm 0.35$ & $1000.41 \pm 0.52$ & $1020.72 \pm 1.33$ &
  & $ 974.36 \pm 0.26$ & $1000.38 \pm 0.41$ &  $1019.72 \pm 0.44$  \\
  19 & $1025.48 \pm 0.63$ & $1049.99 \pm 0.35$ & $\cdots$ &
  &$1025.16\pm0.37$&$1049.34\pm0.29$&$1072.49\pm0.62$\tablenotemark{a}\\
  20 & $1076.70 \pm 0.29$ & $\cdots$ & $\cdots$ &
  & $1076.52 \pm 0.51$ & $\cdots$ & $\cdots$ 
  \enddata
  \tablenotetext{*}{\footnotesize Reference value of $n$, not used for model-fitting.\ \ $^{\rm a}$ Observed mode adopted for refined model-fitting.}
  %\tablenotetext{a}{\footnotesize Observed mode adopted for refined model-fitting.}
  \tablenotetext{b}{\footnotesize Mode not present in any of the optimal models.\ \ $^{\rm c}$ Models suggest an alternate mode identification (see \S\ref{SEC6}).\\}
  %\tablenotetext{c}{\footnotesize Models suggest an alternate mode identification (see \S\ref{SEC6}).}
  \end{deluxetable*}
%-------------------------------------------------------------------------

\section{OSCILLATION FREQUENCIES\label{SEC2}} 

The 58.85-second (short-cadence) photometric data on KIC~11026764 came 
from the first 33.5~d of science operations (2009 May 12 to June 14). Time 
series data were then prepared from the raw observations in the manner 
described by \cite{gil10b}. The power spectrum is shown in Figures 1 and 2 
of \cite{cha10}. Eight teams extracted estimates of the mode frequencies 
of the star. The teams used slightly different strategies to extract those 
estimates, but the main idea was to maximize the likelihood \citep{and90} 
of a multi-parameter model designed to describe the frequency-power 
spectrum of the time series. The model included Lorentzian peaks to 
describe the p-modes, with flat and power-law terms in frequency 
\citep[e.g.,][]{har85} to describe instrumental and stellar background 
noise.

The fitting strategies followed well-established recipes. Some teams 
performed a global fit---optimizing simultaneously every free parameter 
needed to describe the observed spectrum \citep[e.g., 
see][]{app08}---while others fit the spectrum a few modes at a time, an 
approach traditionally adopted for Sun-as-a-star data \citep[e.g., 
see][]{cha99}. Some teams also incorporated a Bayesian approach, with the 
inclusion of priors in the optimization and Markov Chain Monte Carlo 
(MCMC) analysis to map the posterior distributions of the estimated 
frequencies \citep[e.g., see][]{ben09,cam10}.

We then implemented a procedure to select two of the eight sets of 
frequencies, which would subsequently be passed to the modeling teams. Use 
of individual sets---as opposed to some average frequency set---meant that 
the modeling could rely on an easily reproducible set of input 
frequencies, which would not be the case for an average set. We selected a 
\emph{minimal frequency} set to represent the modes that all teams agreed 
upon within the errors, and a \emph{maximal frequency} set, which included 
all possible frequencies identified by at least two of the teams, as 
explained below.

From the sets of frequencies $\nu_{nl,i}$ provided by the eight teams, we 
calculated a list of average frequencies $\bar{\nu}_{nl}$. For each mode 
$\{ n,l \}$, we computed the number of teams returning frequencies that 
satisfied
 \begin{equation}
 \left| \nu_{nl,i} - \bar{\nu}_{nl} \right| \leq \sigma_{nl,i},
 \label{eq:ineq}
 \end{equation} 
with $\sigma_{nl,i}$ representing the frequency uncertainties returned by 
each team. We then compiled a \emph{minimal} list of modes. For each $\{ 
n,l \}$ we counted the total number of teams with identified frequencies, 
as well as the number of those frequencies that satisfied 
Eq.(\ref{eq:ineq}). Modes for which \emph{all} identifications satisfied 
the inequality were added to the minimal list. We also compiled a 
\emph{maximal} list of modes, subject to the much more relaxed criterion 
that the $\{ n,l \}$ satisfying Eq.(\ref{eq:ineq}) should be identified by 
at least two teams.

In the final stage of the procedure, we computed for each of the eight 
frequency sets the normalized root-mean-square ({\it rms}) deviations with 
respect to the $\bar{\nu}_{nl}$ of the minimal and maximal lists of modes. 
The frequency set with the smallest {\it rms} deviation with respect to 
the minimal list was chosen to be the \emph{minimal frequency} set, while 
the set with the smallest {\it rms} deviation with respect to the maximal 
list was chosen to be the \emph{maximal frequency} set. The minimal 
frequency set was also used by \cite{cha10}, and provided the initial 
constraints for the modeling teams (see Table~\ref{tab1}). The maximal 
frequency set was used later for additional validation, as explained in 
\S\ref{SEC5}. Note that the same modes have slightly different frequencies 
in these two sets, since they come from individual analyses. The true radial
order ($n$) of the modes can only be determined from a stellar model, so we
provide arbitrary reference values for convenience. 

%%%%%%%%%%%%%%%%%%%%%%%%%%%%%%%%%%%%%%%%%%%%%%%%%%%%%%%%%%%%%%%%%%%%%%%%%%%

\section{GROUND-BASED DATA\label{SEC3}} %From Molenda-\.Zakowicz et al.

KIC~11026764 ($\alpha_{\rm 2000} = 19^{\rm h}21^{\rm m}24.\!\!^{\rm s}65$, 
$\delta_{\rm 2000} = +48^\circ30^{\prime}53.\!\!^{\prime\prime}2$) has a 
magnitude of $V= 9.55$. The atmospheric parameters given in the Kepler 
Input 
Catalog\footnote{http://archive.stsci.edu/kepler/kepler\_fov/search.php} 
\citep[KIC;][]{lat05} as derived from photometric observations acquired in 
the Sloan filters are $T_{\rm eff}=5502$~K, $\log g = 3.896$~dex, and 
[Fe/H]\,$=-0.255$~dex. The quoted uncertainties on these values are 200~K 
in $T_{\rm eff}$ and 0.5~dex in $\log g$ and [Fe/H]. Since this level of 
precision is minimally useful for asteroseismic modeling, we acquired a 
high-resolution spectrum of the star to derive more accurate values of its 
effective temperature, surface gravity, and metallicity.

\subsection{Observations and Data Reduction\label{sec3.1}}
 
The spectrum was acquired with the Fibre-fed Echelle Spectrograph (FIES) 
at the 2.56-m Nordic Optical Telescope (NOT) on 2009 November 9 (HJD 
2455145.3428). The 1800~s exposure covers the wavelength range 
3730-7360~\AA\ at a resolution R$\sim$67000 and signal-to-noise ratio 
S/N\,$=80$ at 4400~\AA. The Th-Ar reference spectrum was acquired 
immediately after the stellar spectrum. The reduction was performed with 
the FIESTOOL software, which was developed specifically for the FIES 
instrument and performs all of the conventional steps of echelle data 
reduction\footnote{http://www.not.iac.es/instruments/fies/fiestool/FIEStool.html}. 
This includes the subtraction of bias frames, modeling and subtraction of 
scattered light, flat-field correction, extraction of the orders, 
normalization of the spectra (including fringe correction), and wavelength 
calibration.

\subsection{Atmospheric Parameters\label{sec3.2}}

We derived the atmospheric parameters of KIC~11026764 using several 
methods to provide an estimate of the external errors on $T_{\rm eff}$, 
$\log g$, and [Fe/H], which would be used in the asteroseismic modeling. 
The five independent reductions included: the 
VWA\footnote{http://www.hans.bruntt.dk/vwa/} software package 
\citep{bruntt04, bruntt10}, the MOOG\footnote{http://verdi.as.utexas.edu/} 
code \citep{sneden1973}, the 
ARES\footnote{http://www.astro.up.pt/$\sim$sousasag/ares/} code 
\citep{sousa07}, the SYNSPEC method \citep{hubeny88,hubeny95}, and the 
ROTFIT code \citep{frasca03, frasca06}. The principal characteristics of 
the methods employed by each of these codes are described below, and the 
individual results are listed in Table~\ref{tab2}.

\begin{figure*}[t]
  \centerline{\includegraphics[angle=90,width=6.5in]{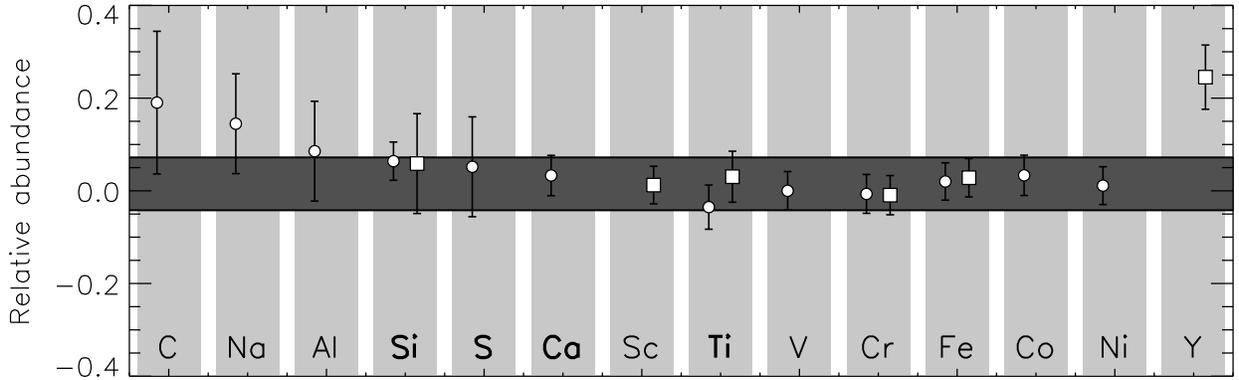}}
  \caption{The mean abundances of the elements in the spectrum of 
  KIC~11026764 as derived with the VWA software, including neutral lines
  (circles) and singly ionized lines (squares). The species labeled in 
  bold are alpha elements. The horizontal bar indicates the mean 
  metallicity of the star and the 1$\sigma$ uncertainty range of the 
  determination. The abundances are given relative to solar
  \citep{grevesse07}.\label{fig2}\\}
  \end{figure*}

For the VWA method, the $T_{\rm eff}$, $\log g$ and microturbulence of the 
adopted MARCS atmospheric models \citep{gus08} are adjusted to minimize 
the correlations of Fe\,I with line strength and excitation potential. The 
atmospheric parameters are then adjusted to ensure agreement between the 
mean abundances of Fe\,I and Fe\,II. Additional constraints on the surface 
gravity come from the two wide Ca lines at 6122 and 6162~\AA, and from the 
Mg-1b lines \citep{bruntt10b}. The final value of $\log g$ is the weighted 
mean of the results obtained from these methods. The mean metallicity is 
calculated only from those elements (Si, Ti, Fe and Ni) exhibiting at 
least 10 lines in the observed spectrum. The uncertainties in the derived 
atmospheric parameters are determined by perturbing the computed models, 
as described in \cite{bruntt08}. Having computed the mean atmospheric 
parameters for the star, VWA finally determines abundances for all of the 
elements contained in the spectrum (see Figure~\ref{fig2}). No trace of 
Li\,I 6707.8 absorption is seen in the spectrum of KIC~11026764. We 
estimate an upper limit for the equivalent width EW\,$\leq 5$~m\AA.

The 2002 version of the MOOG code determines the iron abundance under the 
assumption of local thermodynamic equilibrium (LTE), using a grid of 1D 
model atmospheres by \cite{kurucz93}. The LTE iron abundance was derived 
from the equivalent widths of 65 Fe\,I and 10 Fe\,II lines in the 
4830-6810~\AA\ range, measured with a Gaussian fitting procedure in the 
IRAF\footnote{IRAF is distributed by the National Optical Astronomy 
Observatories, which are operated by the Association of Universities for 
Research in Astronomy, Inc., under cooperative agreement with the National 
Science Foundation.} task {\it splot}. For the analysis, we followed the 
prescription of \cite{randich2006}, using the same list of lines as 
\cite{biazzo10}. The effective temperature and microturbulent velocity 
were determined by requiring that the iron abundance be independent of the 
excitation potentials and the equivalent widths of Fe\,I lines. The 
surface gravity was determined by requiring ionization equilibrium between 
Fe\,I and Fe\,II. The initial values for the effective temperature, 
surface gravity, and microturbulence were chosen to be solar ($T_{\rm eff} 
= 5770$~K, $\log g = 4.44$~dex, and $\xi = 1.1$~km\,s$^{-1}$).

ARES provides an automated measurement of the equivalent widths of 
absorption lines in stellar spectra: the LTE abundance is determined 
differentially relative to the Sun with the help of MOOG and a grid of 
ATLAS-9 plane-parallel model atmospheres \citep{kurucz93}. We used ARES 
with two different lists of iron lines: a `short' one composed of isolated 
iron lines \citep{sousa06}, and a `long' one composed of iron lines 
suitable for automatic measurements \citep{sousa08}. Our computations 
resulted in two consistent sets of atmospheric parameters for 
KIC~11026764.

SYNSPEC provides synthetic spectra based on model atmospheres, either 
calculated by TLUSTY or taken from the literature. We used the new grid of 
ATLAS-9 models \citep{kurucz93,ck03} to calculate synthetic spectra, which 
were then compared to the observed spectrum. Based on the list of iron 
lines from \cite{sousa08}, we derived the stellar parameters in two ways 
to estimate the uncertainty due to the normalization of the observed 
spectrum. For the first approach we determined the minimum $\chi^2$ of the 
deviation between the synthetic iron lines and the observed spectrum for a 
fixed set of stellar parameters. For the second approach, we determined 
the best fitting effective temperature and surface gravity for each iron 
line from the list, and adopted stellar parameters from the mean. The 
primary uncertainty in the final parameters arises from the correlation 
between the effective temperature and surface gravity: a reduction of the 
effective temperature can be compensated by a reduction of the gravity. 
While the $\chi^2$ method results in lower values for both parameters 
($T_{\rm eff}=5560$~K, $\log g=3.62$~dex), the averaging approach yields 
higher values ($T_{\rm eff}=5701$~K, $\log g=3.95$~dex). The differences 
between the two results exceed the formal errors of each method. We 
therefore adopt the mean, and assume half of the difference for the 
uncertainty. The metallicity is determined by minimizing the scatter in 
the parameters derived from individual lines.

% TABLE 2 ----------------------------------------------------------------
  \tablewidth{0pt}
  \tabletypesize{\normalsize}
  \tablecaption{Atmospheric parameter estimates for KIC~11026764.\label{tab2}}
  \begin{deluxetable}{lccc}
  \tablehead{
  \colhead{$T_{\rm eff}$}&\colhead{$\log g$}&\colhead{[Fe/H]}&\colhead{Method}}
  \startdata
  $5640\pm 80$   & $3.84\pm0.10$ & $+0.02\pm0.06$& {\sc vwa}\\
  $5750\pm 50$   & $4.10\pm0.10$ & $+0.11\pm0.06$& {\sc moog}\\
  $5774\pm 39$   & $4.01\pm0.07$ & $+0.09\pm0.03$& {\sc ares}\tablenotemark{a}\\
  $5793\pm 26$   & $4.06\pm0.04$ & $+0.10\pm0.02$& {\sc ares}\tablenotemark{b}\\
  $5630\pm 70$   & $3.79\pm0.17$ & $+0.10\pm0.07$& {\sc synspec}\\
  $5777\pm 77$   & $4.19\pm0.16$ & $+0.07\pm0.08$& {\sc rotfit}
  \enddata
  \tablenotetext{a}{\footnotesize 40 Fe\,I and 12 Fe\,II lines from \cite{sousa06}.}
  \tablenotetext{b}{\footnotesize 247 Fe\,I and 34 Fe\,II lines from \cite{sousa08}.}
  \tablenotetext{}{~}
  \end{deluxetable}
%-------------------------------------------------------------------------

% FIGURE 2 ---------------------------------------------------------------
%-------------------------------------------------------------------------

ROTFIT performs a simultaneous and fast determination of $T_{\rm eff}$, 
$\log g$, and [Fe/H] for a star---as well as its projected rotational 
velocity $v\sin i$---by comparing the observed spectrum with a library of 
spectra for reference stars \citep[see][]{katz98,soubiran98}. The adopted 
estimates for the stellar parameters come from a weighted mean of the 
parameters for the 10 reference stars that most closely resemble the 
target spectrum, which is quantified by a $\chi^2$ measure. We applied the 
ROTFIT code to all echelle orders between 21-69, which cover the range 
4320-6770 \AA\ in the observed spectrum. We also derived a projected 
rotational velocity for KIC~11026764 of $2.8\pm1.6$ km\,s$^{-1}$.

\subsection{Adopted Spectroscopic Constraints\label{sec3.3}}

The effective temperature of KIC~11026764 derived from the five methods 
outlined above generally have overlapping 1$\sigma$ errors. They all point 
to a star that is hotter than the KIC estimate by 100-300\,K. We find 
reasonable agreement between the derived values for surface gravity and 
the KIC estimate. Most of the applied methods result in $\log g$ above 
4.0~dex, again slightly higher than in the KIC. Only VWA and SYNSPEC yield 
slightly lower values, which is not surprising considering the correlation 
between $T_{\rm eff}$ and $\log g$. Finally, all of the methods agree that 
the star is slightly metal-rich compared to the Sun, in contrast to the 
photometric estimate of [Fe/H]\,$=-0.255$~dex from the KIC. The initial 
set of spectroscopic constraints provided to the modeling teams (see 
\S\ref{SEC4}) came from a mean of the preliminary results from the 
analyses discussed above, with uncertainties large enough to cover the 
full range for each parameter: $T_{\rm eff}=5635\pm185$~K, $\log 
g=3.95\pm0.25$~dex, and [Fe/H]\,$=-0.06\pm0.25$~dex.

For the final estimate of the atmospheric parameters (see \S\ref{SEC5}) we 
adopted the results from the VWA method, since it has been carefully 
tested against direct methods for 10 nearby solar-type stars. 
Specifically, \cite{bruntt10} used VWA to determine $T_{\rm eff}$ from 
high-quality spectra and these values were compared to a direct method 
(nearly independent of model atmospheres) using the measured angular 
diameters from interferometry and bolometric flux measurements. A 
comparison of the direct (interferometric) and indirect (VWA) methods 
showed only a slight offset of $-40\pm20$~K, and this offset has been 
removed for KIC~11026764. Similarly, \cite{bruntt10} determined the 
spectroscopic $\log g$ parameter, which agrees very well for the three 
binary stars in their sample where the absolute masses and radii (and 
hence $\log g$) have been measured. Based on these comparisons of direct 
and indirect methods, \cite{bruntt10} also discuss the issue of realistic 
uncertainties on spectroscopic parameters, and their estimates of the 
systematic uncertainties have been incorporated into the adopted values 
from VWA listed in Table~\ref{tab2}.

%%%%%%%%%%%%%%%%%%%%%%%%%%%%%%%%%%%%%%%%%%%%%%%%%%%%%%%%%%%%%%%%%%%%%%%%%%%

\section{STELLAR MODEL SEARCH\label{SEC4}}

Starting with the minimal frequency set described in \S\ref{SEC2} and the 
initial set of spectroscopic constraints from \S\ref{SEC3}, eleven teams 
of modelers performed a ``meta-search'' of the parameter space. Each 
modeler had complete freedom to decide on the input physics and fitting 
strategy to optimize the match to the observations. The results of the 
individual fits were evaluated in a uniform manner and ranked according to 
the total $\chi^2$ between the observed and calculated values of the 
individual oscillation frequencies and the spectroscopic properties. These 
individual fits are listed in Table~\ref{tab3}, and the details of the 
codes and fitting strategies employed by each team of modelers are 
described in the following subsections.

% TABLE 3 ----------------------------------------------------------------
  \tablewidth{0pt}
  \tabletypesize{\normalsize}
  \tablecaption{Initial model-fitting search for KIC~11026764.\label{tab3}}
  \begin{deluxetable*}{lcccccccrcrr}
  \tablehead{
  \colhead{Model}&\colhead{$M/M_\odot$}&\colhead{$Z_{\rm s}$}&\colhead{$Y_{\rm s}$}&
  \colhead{$\alpha$}&\colhead{$t$~(Gyr)}&\colhead{$L/L_\odot$}&\colhead{$R/R_\odot$}&
  \colhead{$T_{\rm eff}$(K)}&\colhead{$\log g$}&\colhead{[Fe/H]}&\colhead{$\chi^2$}}
  \startdata
  A\dotfill&1.13&0.019&0.291&1.70&5.967&3.523&2.026&5562&3.878&$+$0.051&9.8   \\
  B\dotfill&1.23&0.015&0.250&1.80&4.861&4.730&2.097&5885&3.885&$-$0.068&12.4  \\
  C\dotfill&1.24&0.021&0.275&1.79&5.231&4.293&2.093&5750&3.890&$+$0.120&22.5  \\
  D\dotfill&1.31&0.044&0.241&1.42&7.775&2.598&2.139&5016&3.895&$+$0.400&49.7  \\
  E\dotfill&1.22&0.013&0.233&1.83&4.745&4.710&2.080&5899&3.888&$-$0.100&55.8  \\
  F\dotfill&1.10&0.014&0.269&1.85&5.839&3.850&2.015&5706&3.870&$-$0.102&60.6  \\
  G\dotfill&1.20&0.022&0.291&1.88&5.100&3.960&2.160&5663&3.880&$+$0.116&66.8  \\
  H\dotfill&1.10&0.010&0.250&1.75&6.752&3.776&2.012&5677&3.872&$-$0.250&284.3 \\
  I\dotfill&1.27&0.021&0.280&0.50&3.206&3.202&1.792&4854&4.034&$+$0.080&637.3 \\
  \tableline
  J\dotfill&1.12&0.016&0.276&1.90&6.505&3.644&2.026&5593&3.870&$-$0.139&$\cdots$      \\
  $\pm$error&0.14&0.007&0.030&0.78&1.961&0.678&0.091&188&0.023&$\pm$0.166&$\cdots$   \\
  K\dotfill&1.13&0.017&0.283&1.80&6.450&3.610&1.988&5634&3.890&$-$0.044&$\cdots$      \\
  $\pm$error&0.13&0.009&0.009&$\cdots$&1.930&0.770&0.080&161&0.018&$\pm$0.250&$\cdots$
  \enddata
  \tablenotetext{}{~}
  \end{deluxetable*}
%-------------------------------------------------------------------------

\subsection{Model A\label{secA}} %From Dogan et al.

We employed the Aarhus stellar evolution code \citep[ASTEC;][]{jcd08a} for 
stellar evolution computations, and the adiabatic pulsation package 
\citep[ADIPLS;][]{jcd08b} for frequency calculations. The input physics 
for the evolution calculations included the OPAL 2005 equation of state 
\citep{rn02}, OPAL opacity tables \citep{ir96} with low-temperature 
opacities from \cite{Ferguson2005}, and the NACRE nuclear reaction rates 
\citep{angulo99}. Convection was treated according to the mixing-length 
theory of \cite{bv58}. We did not include diffusion or convective 
overshoot in the models.

We computed several grids of evolutionary tracks spanning the parameter 
space around the values given by \cite{cha10}. We primarily adjusted the 
stellar mass and metallicity in our grids, while fixing the mixing-length 
parameter $\alpha$ to 1.7. We scanned the parameter space in mass $M$ from 
1.00 to 1.35 $M_{\odot}$; initial heavy-element mass fraction $Z_{\rm i}$ 
from 0.009 to 0.025; and initial hydrogen mass fraction $X_{\rm i}$ from 
0.68 to 0.76. These values of $Z_{\rm i}$ and $X_{\rm i}$ cover a range of 
$(Z/X)_{\rm i}=0.012$-0.037, or [Fe/H]\,$=-0.317$ to $+$0.177 dex using 
[Fe/H]\,$=\log(Z/X) - \log(Z/X)_\odot$, where $(Z/X)$ is the ratio at the 
stellar surface and $(Z/X)_{\odot}=0.0245$ \citep{GrevesseNoels93}. This 
range of [Fe/H] is compatible with the initial spectroscopic constraints. 
However, we later extended our grids to determine whether there was a 
better model with lower or higher metallicity. For all of the models on 
our tracks, we calculated the oscillation frequencies when the values of 
$T_{\rm eff}$ and $\log g$ were within 2$\sigma$ of the derived values 
(see \S\ref{sec3.3}). We then assigned a goodness of fit to the frequency 
set of each model by calculating $\chi^2$:
 \begin{equation}
 \chi^{2}=\frac{1}{N}\sum_{n,l}\left(\frac{\nu_{l}^{\rm{obs}}(n)-\nu_{l}^{\rm{model}}(n)}{\sigma(\nu_{l}^{\rm{obs}}(n))}\right)^{2},
 \label{chisq}
 \end{equation}
where $N$ is the total number of modes included, $\nu_{l}^{\rm{obs}}(n)$ 
and $\nu_{l}^{\rm{model}}(n)$ are the observed and model frequencies for a 
given spherical degree $l$ and radial order $n$, while 
$\sigma(\nu_{l}^{\rm{obs}}(n))$ represents the uncertainties on the 
observed frequencies. We calculated $\chi^2$ after correcting the 
frequencies for surface effects, following \cite{kje08}:
 \begin{equation}
 \nu_{\rm{obs}}(n)-\nu_{\rm{best}}(n)=a\left[\frac{\nu_{\rm{obs}}(n)}{\nu_{0}}\right]^b,
 \end{equation}
where $\nu_{\rm{obs}}(n)$ and $\nu_{\rm{best}}(n)$ are the observed and 
best model frequencies with spherical degree $l=0$ and radial order $n$, 
and $\nu_{0}$ is the frequency of maximum power in the oscillation 
spectrum, which is 857$\mu$Hz for KIC~11026764. We fixed the exponent $b$ 
to the value derived for the Sun ($b=4.90$) by \cite{kje08}, and $a$ was 
calculated for each model. We computed smaller and more finely sampled 
grids around the models with the lowest $\chi^2$ to refine the fit. The 
properties of the best model are listed in Table~\ref{tab3}. Although we 
found models with higher or lower $\log g$ that had large separations 
quite close to the observed value, the individual frequencies were not 
close to the observations, and the larger range of metallicity did not 
yield improved results. We also found more massive models (around 1.3 
$M_{\odot}$) with a total $\chi^2$ value comparable to our best fit, but 
they did not include the mixed modes. Model A is the best match from the 
family of solutions (also including Models F, H, J and K) with masses near 
1.1 $M_\odot$.

\subsection{Model B\label{secB}} %From Eggenberger et al.

We used the Geneva stellar evolution code including rotation \citep{egg08} 
for all computations. This code includes the OPAL equation of state 
\citep{rn02}, the OPAL opacities \citep{ir96} complemented at low 
temperatures with the molecular opacities of \cite{af94}, the NACRE 
nuclear reaction rates \citep{angulo99} and the standard mixing-length 
formalism for convection \citep{bv58}. Overshooting from the convective 
core into the surrounding radiatively stable layers by a distance 
$d_{\mathrm{ov}} \equiv \alpha_{\mathrm{ov}} \min[H_p,r_{\mathrm{core}}]$ 
\citep{mae89} is included with an overshoot parameter 
$\alpha_{\mathrm{ov}}=0.1$.

In the Geneva code, rotational effects are computed in the framework of 
shellular rotation. The transport of angular momentum then obeys an 
advection-diffusion equation \citep{zah92, mae98}, while the vertical 
transport of chemicals through the combined action of vertical advection 
and strong horizontal diffusion can be described as a purely diffusive 
process \citep{cha92}. Since the modeling of these rotational effects has 
been described in previous papers \citep[e.g.][]{egg10}, we simply note 
that the Geneva code includes a comprehensive treatment of shellular 
rotation and that meridional circulation is treated as a truly advective 
process. For a detailed analysis of the effect of centrifugal force on the 
oscillation frequencies, see Appendix~\ref{APPA}. In addition to rotation, 
atomic diffusion of helium and heavy elements is included with diffusion 
coefficients calculated according to the prescription of \cite{paq86}.

The properties of a stellar model including rotation depend on six 
parameters: the mass $M$, the age $t$, the mixing-length parameter $\alpha 
\equiv l/H_p$ for convection, the initial rotation velocity on the ZAMS 
and two parameters describing the initial chemical composition of the 
star. For these two parameters, we chose the initial helium abundance 
$Y_{\rm i}$ and the initial ratio between the mass fraction of heavy 
elements and hydrogen $(Z/X)_{\rm i}$. This ratio can be related to the 
metallicity [Fe/H] assuming that $\log(Z/X) \cong \mathrm{[Fe/H]} + 
\log(Z/X)_{\odot}$; we adopt the solar value $(Z/X)_{\odot}=0.0245$ given 
by \cite{GrevesseNoels93}. For these computations, the mixing-length 
parameter was fixed to a solar calibrated value ($\alpha_{\odot}=1.7998$) 
and the initial rotation velocity on the ZAMS was 50\,km\,s$^{-1}$. The 
braking law of \cite{kaw88} was used to reproduce the magnetic braking 
experienced by low-mass stars during main-sequence evolution.

With the above assumptions, the characteristics of a stellar model depend 
on only four parameters: $M$, $t$, $Y_{\rm i}$ and $(Z/X)_{\rm i}$. The 
determination of the parameters that best reproduce the observational 
constraints was then performed in two steps as described in \cite{egg06}. 
First, a grid of models with global properties in reasonable agreement 
(within $2\sigma$) with the adopted spectroscopic constraints was 
constructed. Theoretical frequencies of $l \leq 2$ modes in the observed 
range of 590-1100~$\mu$Hz were computed using the adiabatic pulsation code 
\citep{jcd08b} along with the characteristic frequency separations. The 
mean large separation was determined by considering only radial modes. The 
effects of incomplete modeling of the external layers on computed 
frequencies were taken into account using the empirical power law given by 
\cite{kje08}. This correction was applied to theoretical frequencies using 
the solar calibrated value of the exponent ($b=4.90$) and calculating the 
coefficient $a$ for each stellar model.

Using spectroscopic measurements of [Fe/H], $T_{\rm eff}$ and $\log g$ 
together with the observed frequencies, a $\chi^2$ minimization was 
performed to determine the set of model parameters that resulted in the 
best agreement with all observational constraints. The properties of the 
best model are listed in Table~\ref{tab3}. This model correctly reproduces 
the spectroscopic measurements of the surface metallicity and $\log g$, 
but exhibits a slightly higher effective temperature. It is in good 
agreement with the asteroseismic data and in particular with the observed 
deviation of the $l=1$ modes from asymptotic behavior. Model B is the best 
match from the family of solutions (also including Models C, E and G) with 
masses near 1.2 $M_\odot$.

\subsection{Model C\label{secC}} %From Silva Aguirre et al.
 
The Garching Stellar Evolution Code \citep[GARSTEC;][]{ws08} is a 
one-dimensional hydrostatic code which does not include the effects of 
rotation. For the model calculations we used the OPAL equation of state 
\citep{rsi96} complemented with the MHD equation of state at low 
temperatures \citep{hm88}, OPAL opacities for high temperatures 
\citep{ir96} and Ferguson's opacities for low temperatures 
\citep{Ferguson2005}, the \cite{gs98} solar mixture, and the NACRE 
compilation of thermonuclear reaction rates \citep{angulo99}. Mixing is 
performed diffusively in convective regions using the mixing-length theory 
for convection in the formulation from \cite{Kippenhahn1990}, and 
convective overshooting can optionally be implemented as a diffusive 
process with an exponential decay of the convective velocities in the 
radiative zone. The amount of mixing for overshooting depends on an 
efficiency parameter $A$ calibrated with open clusters (typically 
$A=0.016$). Atomic diffusion can be applied following the prescription of 
\cite{thouletal94}, and we use a plane-parallel Eddington grey atmosphere.

We started all of our calculations from the pre-main sequence phase. The 
value of the mixing-length parameter for convection was fixed 
($\alpha=1.791$ from our solar calibration), the Schwarzschild criterion 
for definition of convective boundaries was used, and we did not consider 
convective overshooting or atomic diffusion. We constructed a grid of 
models in the mass range between $1.0\ M_\odot$ and $1.3\ M_\odot$ (in 
steps of 0.01) for several [Fe/H] values from the spectroscopic analysis: 
0.06, 0.09, 0.12, and 0.15. To convert the observed values into total 
metallicity, we applied a chemical enrichment law of $\Delta Y/\Delta Z = 
2$ and used the primordial abundances from our solar calibration. We did 
not explore variations in either the hydrogen or helium abundances.

Once all of the tracks were computed, we restricted our analysis to those 
models contained within the spectroscopic uncertainties. For these cases, 
we calculated the oscillation frequencies using the adiabatic pulsation 
package \citep[ADIPLS;][]{jcd08b} and looked for the model which best 
reproduced the large frequency separation (no surface correction was 
applied to the calculated frequencies). Several models were found to 
fulfill these requirements for each metallicity grid, and among those best 
fit models to the large frequency separation we then performed a $\chi^2$ 
test to obtain the global best fit to the individual frequencies and the 
spectroscopic constraints. Our global best fit model came from the grid 
with [Fe/H]\,$=0.12$, and the properties are listed in Table~\ref{tab3}. 
This model reproduces well the observed mixed modes.

\subsection{Model D\label{secD}} %From Metcalfe et al.

The Asteroseismic Modeling Portal (AMP) is a web-based tool tied to 
TeraGrid computing resources that uses the Aarhus stellar evolution code 
\citep[ASTEC;][]{jcd08a} and adiabatic pulsation code 
\citep[ADIPLS;][]{jcd08b} in conjunction with a parallel genetic algorithm 
\citep{mc03} to optimize the match to observational data 
\citep[see][]{mcc09}. The models use the OPAL 2005 equation of state 
\citep[see][]{rn02} and the most recent OPAL opacities 
\citep[see][]{ir96}, supplemented by Kurucz opacities at low temperatures. 
The nuclear reaction rates come from \cite{bp95}, convection is described 
by the mixing-length theory of \cite{bv58}, and we can optionally include 
the effects of helium settling as described by \cite{mp93}.

Each model evaluation involves the computation of a stellar evolution 
track from the zero-age main sequence (ZAMS) through a mass-dependent 
number of internal time steps, terminating prior to the beginning of the 
red giant stage. Rather than calculate the pulsation frequencies for each 
of the 200-300 models along the track, we exploit the fact that the 
average frequency separation of consecutive radial orders 
$\left<\Delta\nu_0\right>$ in most cases is a monotonically decreasing 
function of age \citep{jcd93}. Once the evolution track is complete, we 
start with a pulsation analysis of the model at the middle time step and 
then use a binary decision tree---comparing the observed and calculated 
values of $\left<\Delta\nu_0\right>$---to select older or younger models 
along the track. This allows us to interpolate the age between the two 
nearest time steps by running the pulsation code on just 8 models from 
each stellar evolution track. The frequencies of each model are then 
corrected for surface effects following the prescription of \cite{kje08}.

The genetic algorithm (GA) optimizes four adjustable model parameters, 
including the stellar mass ($M$) from 0.75 to 1.75 $M_\odot$, the 
metallicity ($Z$) from 0.002 to 0.05 (equally spaced in $\log Z$), the 
initial helium mass fraction ($Y_{\rm i}$) from 0.22 to 0.32, and the 
mixing-length parameter ($\alpha$) from 1 to 3. The stellar age ($t$) is 
optimized internally during each model evaluation by matching the observed 
value of $\left<\Delta\nu_0\right>$ (see above). The GA uses two-digit 
decimal encoding, such that there are 100 possible values for each 
parameter within the specified ranges. Each run of the GA evolves a 
population of 128 models through 200 generations to find the optimal set 
of parameters, and we execute 4 independent runs with different random 
initialization to ensure that the best model identified is truly the 
global solution. The resulting properties of the optimal model are listed 
in Table~\ref{tab3}.

The extreme values in this global fit arose from treating each 
spectroscopic constraint as equivalent to a single frequency. Since the 
adopted spectroscopic errors are large, they provide much more flexibility 
for the models compared to the individual frequencies with relatively 
small errors. Consequently, the fitting algorithm found it advantageous to 
shift the effective temperature and metallicity by several $\sigma$ from 
their target values to achieve significantly better agreement with the 22 
oscillation frequencies. The improvement in the frequency match outweighed 
the degradation in the spectroscopic fit for the calculation of $\chi^2$. 
The solution to this problem may be to calculate a separate value of 
$\chi^2$ for the asteroseismic and spectroscopic constraints, and then 
average them to provide more equal weight to the two types of constraints.
This is an important lesson for future automated searches, and explains 
why Model D does not align with either of the two major families of 
solutions.

\subsection{Model E\label{secE}} %From Basu et al.

We used the Yale Stellar Evolution Code \citep[YREC;][]{demarque2008} in 
its non-rotating configuration to model KIC~11026764. All models were 
constructed with the OPAL equation of state \citep{rn02}. We used OPAL 
high temperature opacities \citep{ir96} supplemented with low temperature 
opacities from \cite{Ferguson2005}. The NACRE nuclear reaction rates 
\citep{angulo99} were used. We assumed that the current solar metallicity 
is that given by \cite{gs98}. We have not explored the consequences of 
using the lower metallicity measurements of \cite{ags05} or the 
intermediate metallicity measurements of \cite{ludwig09}. We searched for 
the best fit within a fixed grid of models. There were eight separate 
grids defined by different combinations of the mixing-length parameter 
($\alpha=1.83$ or $\alpha=2.14$), the initial helium abundance (either 
$Y_{\rm i}=0.27$ or $Y_{\rm i}$ calculated assuming a $\Delta Y/\Delta 
Z=2$, with $Y_{\rm i}$ for [Fe/H]\,$=0$ being the current solar CZ helium 
abundance), and the amount of overshoot ($0\,H_p$ or $0.2\,H_p$). All 
models included gravitational settling of helium and heavy elements using 
the formulation of \cite{thouletal94}.

Our fitting method included two steps. In the first step, we calculated 
the average large frequency separation for the models and selected all of 
those that fit the observed separation within 3$\sigma$ errors. We adopted 
the observed value of $\Delta\nu=50.8\pm 0.3\mu$Hz from \cite{cha10}. A 
second cut was made using the effective temperature: all models within 
$\pm 2\sigma$ of the observed value were chosen. A third cut was made 
using the frequencies of the three lowest-frequency $l=0$ modes. Given the 
small variation in mass, this process was effectively a radius cut. The 
selected models had radii around 2$R_\odot$. Note that the Yale-Birmingham 
radius pipeline \citep{Basu10} finds a radius of 
$2.18^{+0.04}_{-0.05}R_\odot$ for this star using the adopted values of 
$\Delta\nu$, $T_{\rm eff}$, $\log g$ and [Fe/H]. In the second step of the 
process, we made a finer grid in mass and age around the selected values 
and then compared the models with the observed set of frequencies. The 
properties of our best fit model are listed in Table~\ref{tab3}. This 
model was constructed with $Y_{\rm i}=0.27$, $Z_{\rm i}=0.0147$ and core 
overshoot of $0.2\,H_p$. We were unable to find a good model without core 
overshoot.

\subsection{Model F\label{secF}} %From Bonanno et al.

We modeled KIC~11026764 with the Catania Astrophysical Observatory version 
of the GARSTEC code \citep{alfio} using a grid-based approach. The input 
physics of this stellar evolution code included the OPAL 2005 equation of 
state \citep{rn02} and the OPAL opacities \citep{ir96} complemented in the 
low temperature regime with the tables of \cite{af94}. The nuclear 
reaction rates were taken from the NACRE collaboration \citep{angulo99} 
and the standard mixing-length formalism for convection was used 
\citep{Kippenhahn1990}. Microscopic diffusion of hydrogen, helium and all 
of the major metals can optionally be taken into account. The outer 
boundary conditions were determined by assuming an Eddington grey 
atmosphere.

A grid of evolutionary models was computed to span the 1$\sigma$ 
uncertainties in the spectroscopic constraints obtained from ground based 
observations. When a given evolutionary track was in the error box, a 
maximum time step of 20 Myr was chosen and the frequencies were computed 
with the ADIPLS code. A non-uniform grid of mass in the range 1.0-$1.24 \; 
M_\odot$, helium abundance in the range $Y_{\rm i}=0.26$-0.31, 
mixing-length parameter $\alpha=1.6$-1.9 and initial surface heavy-element 
abundances $(Z/X)_{\rm i}=0.022$-0.029 was scanned. A global optimization 
strategy was implemented by minimizing the $\chi^2$ for all of the $l=0$, 
$l=1$ and $l=2$ modes. The empirical surface effect, as discussed by 
\cite{kje08}, was used to correct all theoretical frequencies. The 
properties of the best model with heavy-element diffusion and surface 
corrected frequencies are listed in Table~\ref{tab3}.

\subsection{Model G\label{secG}} %From DiMauro et al.

We used a version of the Aarhus stellar evolution code 
\citep[ASTEC;][]{jcd08a} which includes the OPAL 2001 equation of state 
\citep{rsi96}, OPAL opacities \citep{ir96}, \cite{bp95} nuclear cross 
sections and the mixing-length formalism \citep{bv58} for convection. We 
computed several grids of models by varying all of the input parameters 
within the range of the observed errors \citep{cha10}. In particular we 
calculated evolutionary tracks by varying the input mass in the range 
$M=0.9$-1.2~$M_{\odot}$, the metallicity in the range $Z=0.009$-0.03, and 
the hydrogen abundance in the range $X=0.67$-0.7. We also adopted 
different values of the mixing-length parameter in the range 
$\alpha=1.67$-1.88. We calculated additional evolutionary models using the 
\cite{cm92} convection formulation. The mixing-length parameter $\alpha$ 
of the CM formulation was chosen in the range $\alpha=\alpha_{\rm 
CM}=0.9$-1.0. To obtain the deviation from asymptotic behavior observed in 
the $l = 1$ modes (the mixed modes) we did not include overshooting in the 
calculation, following the conclusion of \cite{dima03}. 
These models are distinct from the grid used to produce Model A, not only
because they employ a slightly older EOS and nuclear reaction rates, but 
also because the grid search included $\alpha$ and calculated fewer models 
within the specified range of parameter values. We used the 
adiabatic oscillation code \citep[ADIPLS;][]{jcd08b} to calculate the 
p-mode eigenfrequencies with harmonic degree $l=0$-2. The characteristics 
of the model which best fits the observations are listed in 
Table~\ref{tab3}.

\subsection{Model H\label{secH}} %From Kosovichev and Kitiashvili
 
For the evolution calculations we used the publicly available Dartmouth 
stellar evolution code 
\citep[DSEP;][]{Chaboyer2001,Guenther1992,Dotter2007}, which is based on 
the code developed by Pierre Demarque and his students 
\citep{Larson1964,Demarque1971}. The input physics includes high 
temperature opacities from OPAL \citep{ir96}, low temperature opacities 
from \cite{Ferguson2005}, the nuclear reaction rates of 
\cite{Bahcall1992}, helium and heavy-element settling and diffusion 
\citep{mp93}, and Debye-H\"uckel corrections to the equation of state 
\citep{Guenther1992}. The models employ the standard mixing-length theory. 
Convective core overshoot is calculated assuming that the extent is 
proportional to the pressure scale height at the boundary 
\citep{Demarque2004}. The models used the standard conversion from [Fe/H] 
and $\Delta Y/\Delta Z$ to $Z$ and $Y$ \citep{Chaboyer1999}. The 
oscillation frequencies were computed using the adiabatic oscillation 
codes of \cite{Kosovichev1999} and \cite{jcd08b}. No surface corrections 
were applied.
 
The strategy to find a model matching the observed spectroscopic 
constraints involved calculating a series of evolutionary tracks in the 
$\log g$-$T_{\rm eff}$ plane for a mass range of 1.0-$1.3~M_\odot$, 
heavy-element abundance $Z=0.01$-0.03, initial helium abundance $Y_{\rm 
i}=0.25$-0.30, and mixing-length parameter $\alpha=1.70$-1.75. We then 
selected the models closest to the target values within half of the 
specified uncertainties. All models for the search were calculated 
assuming an overshoot parameter $\alpha_{\rm ov}=0.2$, and included 
element diffusion. For comparison, the corresponding models without 
diffusion and convective overshoot were also calculated. The oscillation 
spectra were matched to the observed frequencies, first by comparing the 
frequencies of radial ($l=0$) modes with the corresponding observed 
frequencies, and then selecting a model with the closest frequency values 
for the $l=1$ and $l=2$ modes. The properties of the final model are 
listed in Table~\ref{tab3}. This model matches the observed frequencies 
quite well except for the first two $l=1$ modes, which deviate by about 
12-13 $\mu$Hz. Our search demonstrated that the behavior of the mixed mode 
frequencies is sensitive to element diffusion and convective overshoot. 
This requires further investigation.

\subsection{Model I\label{secI}} %From Suarez et al.
 
To characterize KIC~11026764 we constructed a grid of stellar models with 
the CESAM code \citep{Morel97}, and computed their oscillation frequencies 
with the adiabatic oscillation code FILOU \citep{SuaThesis, Sua08filou}. 
Opacity tables were taken from the OPAL package \citep{ir96}, complemented 
at low temperatures ($T\leq10^4\,K$) with the tables provided by 
\cite{af94}. The atmosphere was constructed from a Eddington~$T$-$\tau$ 
relation and was assumed to be grey. The stellar metallicity $(Z/X)$ was 
derived from the [Fe/H] value assuming $(Z/X)_\odot=0.0245$ 
\citep{GrevesseNoels93}, $Y_{\mathrm{pr}}=0.235$ and $Z_{\mathrm{pr}}=0$ 
for the primordial helium and heavy-element abundances, and a value 
$\Delta Y/\Delta Z=2$ for the enrichment ratio. No microscopic diffusion 
of elements was considered.

The main strategy was to search for representative equilibrium models of 
the star in a database of $5\times10^5$ equilibrium models, querying for 
those matching the global properties of the star, including the effective 
temperature, gravity and metallicity. Using this set of models, we then 
applied the asteroseismic constraints, including the individual 
frequencies and large separations. The global fitting method involved a 
$\chi^2$ minimization, taking into account all of the observational 
constraints simultaneously. No correction for surface effects was applied, 
and no {\it a priori} information on mode identification was assumed when 
fitting the individual frequencies. The properties of the best model we 
found is listed in Table~\ref{tab3}, and includes overshooting with 
$\alpha_{\rm ov}=0.3$. Note that the analysis did not adopt the 
identifications of spherical degree ($l$) from \S\ref{SEC2}, and it 
included $l=3$ modes to perform the match. As a consequence, the final 
result is much different than any of the others and it does not fall into 
either of the two major families of solutions.

\subsection{Model J\label{secJ}} %From Quiron et al.

The SEEK procedure makes use of a large grid of stellar models computed 
with the Aarhus stellar evolution code \citep[ASTEC;][]{jcd08a}. It 
compares the observations with every model in the grid and makes a 
probabilistic assessment, with the help of Bayesian statistics, about the 
global properties of the star. The model grid includes 7,300 evolution 
tracks containing 5,842,619 individual models. Each track begins at the 
ZAMS and continues to the red giant branch or a maximum age of $t = 
15$~Gyr. The tracks are separated into 100 subsets with different 
combinations of metallicity $Z$, initial hydrogen content $X_{\rm i}$ and 
mixing-length parameter $\alpha$. These combinations are separated into 
two regularly spaced and interlaced subgrids. The first subgrid comprises 
tracks with $Z = [0.005, 0.01, 0.015, 0.02, 0.025, 0.03]$, $X_{\rm i} = 
[0.68, 0.70, 0.72, 0.74]$, and $\alpha=[0.8, 1.8, 2.8]$ while the second 
subset has $Z = [0.0075, 0.0125, 0.0175, 0.0225, 0.0275]$, $X_{\rm i} = 
[0.69, 0.71, 0.73]$, $\alpha=[1.3, 2.3]$. Every subset is composed of 73 
tracks with masses between 0.6 and 3.0~$M_{\rm \odot}$. The spacing in 
mass between the tracks is 0.02 $M_{\rm \odot}$ from 0.6 to 1.8 $M_{\rm 
\odot}$ and 0.1 from 1.8 to 3.0 $M_{\rm \odot}$. A relatively high value 
of $Y_\odot = 0.2713$ and $Z_\odot = 0.0196$ for the Sun has been used for 
the standard definition of [Fe/H] in SEEK. This value is used to calibrate 
solar models from ASTEC to the correct luminosity 
\citep{Christensen-Dalsgaard98}. The input physics include the OPAL 
equation of state, opacity tables from OPAL \citep{ir96} and \cite{af94}, 
and the metallic mixture of \cite{gs98}. Convection is treated according 
to the mixing-length theory of \cite{bv58} with the convective efficiency 
characterized by the mixing-length to pressure height scale ratio 
$\alpha$, which varies across the grid of models. Diffusion and 
overshooting were not included.

% TABLE 4 ----------------------------------------------------------------
  \tablewidth{0pt}
  \tabletypesize{\normalsize}
  \tablecaption{Final model-fitting results for KIC~11026764.\label{tab4}}
  \begin{deluxetable*}{lcccccccrcrr}
  \tablehead{
  \colhead{Model}&\colhead{$M/M_\odot$}&\colhead{$Z_{\rm s}$}&\colhead{$Y_{\rm s}$}&
  \colhead{$\alpha$}&\colhead{$t$~(Gyr)}&\colhead{$L/L_\odot$}&\colhead{$R/R_\odot$}&
  \colhead{$T_{\rm eff}$(K)}&\colhead{$\log g$}&\colhead{[Fe/H]}&\colhead{$\chi^2$}}
  \startdata
  FA\dotfill&1.13&0.017&0.305&1.64&5.268&4.141&2.036&5778&3.872&$+$0.009&3.69         \\
  AA\dotfill&1.13&0.019&0.291&1.70&5.977&3.520&2.029&5556&3.877&$+$0.051&6.11         \\
  AA$^\prime$\dotfill&1.13&0.019&0.291&1.70&5.935&3.454&2.026&5534&3.877&$+$0.031&7.40\\
  GA\dotfill&1.10&0.017&0.296&1.88&6.100&3.420&2.010&5539&3.870&$+$0.004&78.05        \\
  CA\dotfill&1.13&0.019&0.291&1.70&6.204&3.493&2.030&5546&3.876&$+$0.050&152.91       \\
  EA\dotfill&1.12&0.019&0.291&1.70&6.683&3.202&2.029&5424&3.870&$+$0.050&230.58       \\
  \tableline
  AB\dotfill&1.23&0.018&0.242&1.80&5.869&3.804&2.083&5591&3.890&$-$0.010&6.97         \\
  AB$^\prime$\dotfill&1.20&0.024&0.276&1.80&5.994&3.460&2.072&5475&3.884&$+$0.146&7.26\\
  BB\dotfill&1.22&0.021&0.270&1.80&5.153&4.190&2.061&5758&3.896&$+$0.072&7.57         \\
  FB\dotfill&1.24&0.021&0.280&1.79&4.993&4.438&2.092&5800&3.890&$+$0.091&8.52         \\
  EB\dotfill&1.22&0.013&0.232&1.80&4.785&4.651&2.079&5882&3.890&$-$0.130&18.54        \\
  CB\dotfill&1.24&0.015&0.250&1.80&5.064&4.696&2.089&5887&3.892&$-$0.080&45.84        \\
  \tableline
  J$^\prime$\dotfill&1.27&0.021&0.270&1.52&4.260&4.011&2.105&5634&3.892&$+$0.080&$\cdots$\\
  $\pm$error&0.09&0.003&0.024&0.74&1.220&0.371&0.064&81&0.020&$\pm$0.060&$\cdots$        \\
  K$^\prime$\dotfill&1.20&0.022&0.278&1.80&5.980&3.700&2.026&5619&3.900&$+$0.070&$\cdots$\\
  $\pm$error&0.04&0.003&0.003&$\cdots$&0.610&0.300&0.027&79&0.006&$\pm$0.060&$\cdots$ 
  \enddata
  \tablenotetext{}{~}
  \end{deluxetable*}
%-------------------------------------------------------------------------

The grid allows us to map the physical input parameters of the model 
$\mathbf{p} \equiv \{M,t,Z,X_{\rm i},\alpha\}$ into the grid of observable 
quantities $\mathbf{q}^\mathbf{g} \equiv \{\Delta\nu,\delta\nu,T_{\rm 
eff},\log g,{\rm [Fe/H]},...\}$, defining the transformation
 \begin{equation}\label{grid}
 \mathbf{q}^\mathbf{g} = \mathcal{K}(\mathbf{p}).
 \end{equation}
We compare these quantities to the observed values 
$\mathbf{q}^\mathbf{obs}$ with the help of a likelihood function 
$\mathcal{L}$,
 \begin{equation}\label{likely}
 \mathcal{L} = \left({\displaystyle\prod_{i=0}^n}\frac{1}{\sqrt{2\pi}\sigma_i}\right)\exp(-\chi^2/2),
 \end{equation}
and the usual $\chi^2$ definition
 \begin{equation}\label{chi2}
 \chi^2 = \frac{1}{N}{\displaystyle\sum_{i=0}^{N}}\left( \frac{{q}_i^{\rm obs} - {q}_i^{\rm g}}{\sigma_i} \right)^2
 \end{equation}
where $\sigma_i$ is the estimated error for each observation $q_i^{\rm 
obs}$, and $N$ is the number of observables. The maximum likelihood is 
then combined with the prior probability of the grid $f_0$ to yield the 
posterior, or the resulting probability density
 \begin{equation}\label{baye_dense}
 f(\mathbf{p}) \propto f_0(\mathbf{p}) \mathcal{L}(\mathcal{K}(\mathbf{p})).
 \end{equation}
This probability density can be integrated to obtain the value and 
uncertainty for each of the parameters, as listed in Tables~\ref{tab3} and 
\ref{tab4}. It can also be projected onto any plane to get the correlation 
between two parameters, as shown in Appendix~\ref{APPB}. The details of 
the SEEK procedure, including the choice of priors, and an introduction to 
Bayesian statistics can be found in \cite{QCDA10}.

SEEK uses the large and small separations and the median frequency of the 
observed modes as asteroseismic inputs. Values of $\Delta \nu_{0} = 50.68 
\pm 1.30$ (computed with $l=0$ modes only) and $\delta \nu_{0,2} = 4.28 
\pm 0.73 $ (computed from $l=0,2$ modes) were derived from {\it Kepler} 
data around a central value 900~$\mu$Hz. Each separation is the mean of 
the individual observed separations, while the error is the standard 
deviation of the individual values from the mean. These values differ
slightly from those given in \cite{cha10} because they are calculated from 
the individual frequencies rather than derived from the power spectrum.
Using these asteroseismic inputs along with the initial spectroscopic 
constraints, we obtained the parameters listed in Table~\ref{tab3}. For a 
SEEK analysis of the importance of the asteroseismic constraints, see 
Appendix~\ref{APPB}.

\subsection{Model K\label{secK}} %From Stello et al.

To investigate how well we can find an appropriate model without comparing 
individual oscillation frequencies, we used the RADIUS pipeline 
\citep{Stello09a}, which takes $T_{\mathrm{eff}}$, $\log g$, [Fe/H], and 
$\Delta\nu$ as the only inputs to find the best fitting model. The value 
of $\Delta\nu=50.8\pm0.3\,\mu$Hz from \cite{cha10} was adopted. The 
pipeline is based on a large grid of ASTEC models \citep{jcd08a} using the 
EFF equation of state \citep{Eggleton73}. We use the opacity tables of 
\cite{RogersIglesias95} and \cite{Kurucz91} for $T<10^4$\,K with the solar 
mixture of \cite{GrevesseNoels93}. Rotation, overshooting and diffusion 
were not included. The grid was created with fixed values of the 
mixing-length parameter ($\alpha=1.8$) and the initial hydrogen abundance 
($X_{\rm i}=0.7$). The resolution in $\log Z$ was 0.1 dex between 
$0.001<Z<0.055$, and the resolution in mass was 0.01 M$_{\odot}$ from 0.5 
to 4.0 M$_{\odot}$. The evolution begins at the ZAMS and continues to the 
tip of the red giant branch. To convert between the model values of $Z$ 
and the observed [Fe/H], we used $Z_\odot=0.0188$ \citep{Cox00}.

We made slight modifications to the RADIUS approach described by 
\cite{Stello09a}. First, the large frequency separation was derived by 
scaling the solar value \citep{KjeldsenBedding95} instead of calculating 
it directly from the model frequencies. Although there is a known 
systematic difference between these two ways of deriving $\Delta\nu$, the 
effect is probably below the 1\% level \citep{Stello09b,Basu10}. Second, 
we pinpointed a single best-fitting model based on a $\chi^2$ formalism 
that was applied to all models within $\pm3\sigma$ of the observed 
properties. The properties of the best fitting model are listed in 
Table~\ref{tab3}. This model shows a frequency pattern in the \'echelle 
diagram that looks very similar to the observations if we allow a small 
tweaking of the adopted frequency separation (52.1$\mu$Hz) used to 
generate the \'echelle. This basically means that we found a model that 
homologously represents the observations quite well. In particular, we see 
relative positions of the $l=1$ mode frequencies that are very similar to 
those observed.\\

%%%%%%%%%%%%%%%%%%%%%%%%%%%%%%%%%%%%%%%%%%%%%%%%%%%%%%%%%%%%%%%%%%%%%%%%%%%

\section{MODEL-FITTING RESULTS\label{SEC5}}

Based on the value of $\chi^2$ from the initial search in \S\ref{SEC4}, we 
adopted two reference models (Models A and B in Table~\ref{tab3}) each 
constructed with a very different set of input physics, but almost equally 
capable of providing a good match to the observations. Note that Model A 
does not include overshoot or diffusion, while Model B includes overshoot, 
diffusion and a full treatment of rotation. These two models differ 
significantly in the optimal values of the mass, effective temperature, 
metallicity and luminosity, but they both agree with the observational 
constraints at approximately the same level ($\chi^2\sim10$). With the 
exceptions of Models D and I (see subsections above), the other 
independent analyses generally fall into the two broad families of 
solutions defined by Models A and B. The lower mass family includes Models 
A, F, H, J and K, while the higher mass family includes Models B, C, E and 
G. We identified several additional asteroseismic constraints from the 
maximal frequency set (see \S\ref{SEC2}) and we adopted revised 
spectroscopic constraints from VWA (see \S\ref{SEC3}) to refine our 
analysis of Models A and B using several different codes.

\subsection{Refining the Best Models\label{sec5.1}}

Comparing the theoretical frequencies of Models A and B with the maximal 
frequency set from \S\ref{SEC2}, we identified four of the seven 
additional oscillation modes that could be used for refined model-fitting 
(see Table~\ref{tab1}). Recall that the maximal frequency set comes from 
the {\it individual analysis} with the smallest {\it rms} deviation with 
respect to the maximal list, so the frequencies of the modes from the 
minimal set are slightly different in the maximal set. Without any 
additional fitting, these subtle frequency differences improve the 
$\chi^2$ of Models A and B when comparing them to those modes from the 
maximal set that are also present in the minimal set. There is one 
additional $l=0$ mode ($n=12$) and three additional $l=2$ modes 
($n=11,16,19$) in the maximal set that are within 3$\sigma$ of frequencies 
in both Models A and B. Considering the very different input physics of 
these two models, we took this agreement as evidence of the reliability of 
these four additional frequencies and we incorporated them as constraints 
for our refined model-fitting. Two of the remaining frequencies in the 
maximal set ($n=12, l=1$ and 2) were not present in either Models A or B, 
while one ($n=10, l=2$) had a close match in Model B but not in Model A. 
We excluded these three modes from the refined model-fitting. Given that 
the $l=1$ modes provide the strongest constraints on the models (see 
\S\ref{sec5.2}), the additional $l=0$ and $l=2$ modes are expected to 
perturb the final fit only slightly.

In addition to the 26 oscillation frequencies from the maximal set, we 
also included stronger spectroscopic constraints in the refined 
model-fitting by adopting the results of the VWA analysis instead of using 
the mean atmospheric parameters from the preliminary analyses (see 
\S\ref{sec3.3}). Although the uncertainties on all three parameters are 
considerably smaller from the VWA analysis, the actual values only differ 
slightly from the initial spectroscopic constraints. These were just three 
of the 25 constraints used to calculate the $\chi^2$ and rank the initial 
search results in Table~\ref{tab3}. Since the 22 frequencies from the 
minimal set were orders of magnitude more precise, they dominated the 
$\chi^2$ determination. Although the spectroscopic constraints from VWA 
are more precise than the initial atmospheric parameters, they are still 
much less precise than the frequencies and should only perturb the 
$\chi^2$ ranking slightly. Consequently, we do not need to perform a new 
global search after adopting the additional and updated observational 
constraints.

Several modeling teams used the updated asteroseismic and spectroscopic 
constraints for refined model-fitting with a variety of codes. Each team 
started with the parameters of Models A and B from Table~\ref{tab3}, and 
then performed a local optimization to produce the best match to the 
observations within each family of solutions. The results of this analysis 
are shown in Table~\ref{tab4}, where the refined Models A and B are ranked 
separately by their final $\chi^2$ value. Each model is labeled with a 
letter from Table~\ref{tab3} to identify the modeling team, followed by 
either A or B to identify the family of solutions. The two pipeline 
approaches labeled J$^\prime$ and K$^\prime$ simply adopted the revised 
constraints to evaluate any shift in the optimal parameter estimates and 
errors. Note that both the SEEK and RADIUS pipelines identified parameters 
in the high-mass family of solutions when using the revised constraints, 
but the low-mass family was only marginally suboptimal. The apparent 
bifurcation of results in Table~\ref{tab4} into two values of $\alpha$ 
arises from the decision of most modelers to fix this parameter in each 
case to the original value from Table~\ref{tab3}. For each family of 
solutions, the modeling teams adopted the appropriate input physics: 
neglecting overshoot and diffusion for the refined Models A, while 
including both ingredients for the refined Models B. One team produced two 
additional models (labeled AA$^\prime$ and AB$^\prime$) to isolate the 
effect of input physics on the final results. Model AA$^\prime$ started 
from the parameters of Model A but included overshoot and helium settling, 
while Model AB$^\prime$ searched in the region of Model B but neglected 
overshoot and diffusion.

% FIGURE 3 ---------------------------------------------------------------
  \begin{figure}[t]
  \centerline{\includegraphics[angle=0,width=\linewidth]{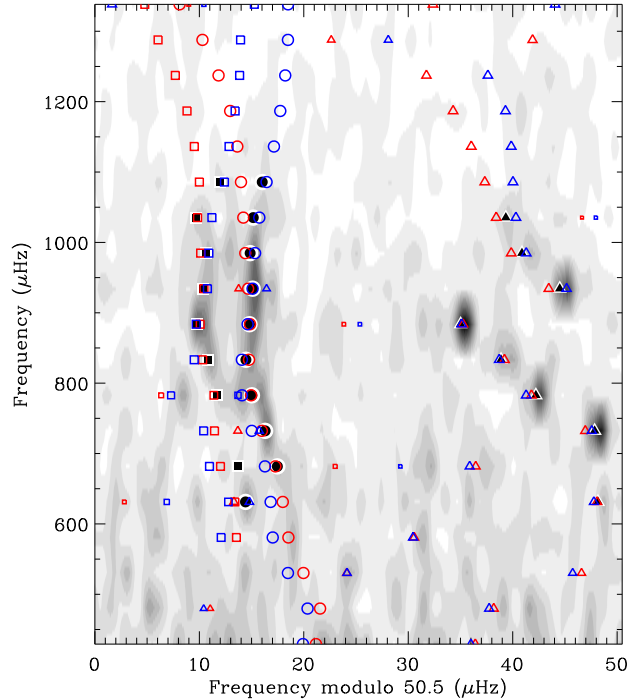}}
  \caption{An \'echelle diagram showing the 26 frequencies from the maximal 
  set that were used as constraints (solid points) with the frequencies of 
  Models AA (blue) and AB (red) for comparison. The star exhibits modes with 
  $l=0$ (circles), $l=1$ (triangles), and $l=2$ (squares). A greyscale map 
  showing the power spectrum (smoothed to 1\,$\mu$Hz resolution) is 
  included in the background for reference.\label{fig3}}
  \end{figure}
%-------------------------------------------------------------------------

\subsection{Stellar Properties \& Error Analysis\label{sec5.2}}

% FIGURE 4 ---------------------------------------------------------------
  \begin{figure*}[t]
  \centerline{\includegraphics[angle=0,width=6.5in]{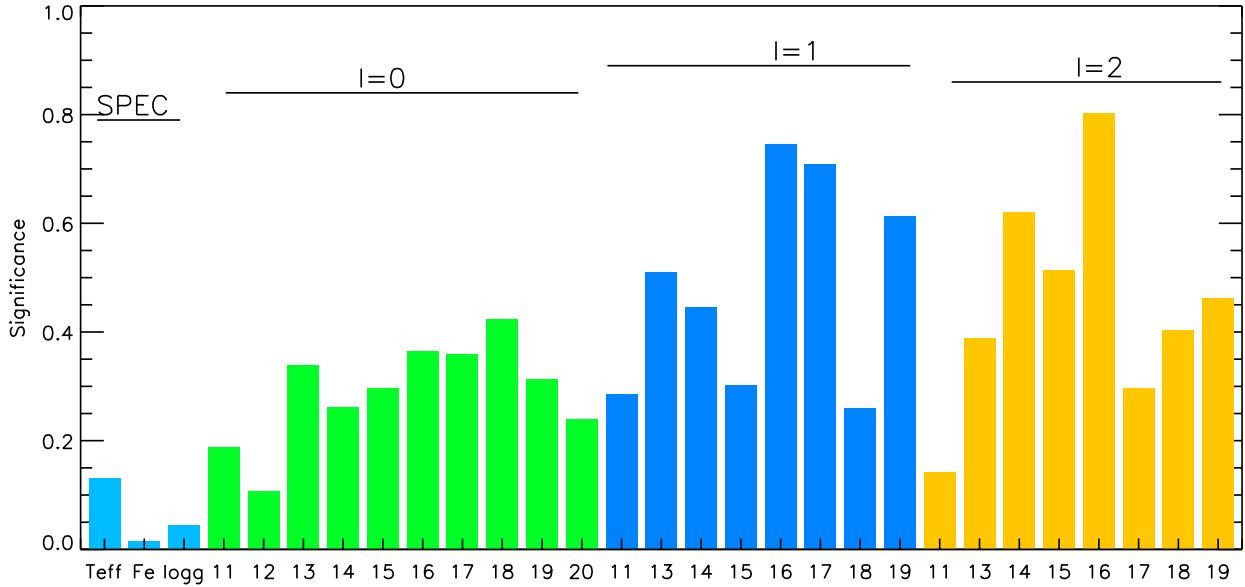}}
  \caption{The significance $S_i$ of each observational constraint in 
  determining the parameters of Model AA. Each oscillation frequency is 
  labeled with the reference radial order from Table~\ref{tab1}, and the 
  degrees $l = 0$, 1 and 2 are shown sequentially in different shades 
  from left to right.\label{fig4}\\}
  \end{figure*}
%-------------------------------------------------------------------------

An inspection of the results in Table~\ref{tab4} reveals that models in 
either family of solutions can provide a comparable match to the 
observational constraints. This ambiguity cannot be attributed to the 
input physics, since the models that sample all four combinations of the 
input physics and family of solutions (AA, AA$^\prime$, AB, AB$^\prime$) 
have comparable $\chi^2$ values. The individual frequencies of these 
models all provide a good fit to the data, including the $l=1$ mixed 
modes. The observations are compared to two representative models in 
Figure~\ref{fig3} using an \'echelle diagram, where we divide the 
frequency spectrum into segments of length $\left<\Delta\nu\right>$ and 
plot them against the oscillation frequency. This representation of the 
data aligns modes with the same spherical degree into roughly vertical 
columns, with $l=0$ modes shown as circles, $l=1$ modes shown as 
triangles, and $l=2$ modes shown as squares. The 26 modes from the maximal 
frequency set that were included in the final fit are indicated with solid 
points. Open points indicate the model frequencies, with Model AA shown in 
blue and Model AB shown in red. A greyscale map of the power spectrum 
(smoothed to 1\,$\mu$Hz resolution) is included in the background for 
reference. The different models appear to be sampling comparable local 
minima in a correlated parameter space. Without additional constraints, we 
have no way of selecting one of these models over the other.

We can understand the two families of models by considering the general 
properties of subgiant stars, where a wide range of masses can have the 
same stellar luminosity with minor adjustments to the input physics and 
other model parameters---in particular the helium mass fraction. This 
degeneracy between mass and helium abundance has been discussed in the 
modeling of specific subgiant stars \citep[e.g.][]{Fernandes03, 
Pinheiro10,Yang10}, and it adds a large uncertainty to the already 
difficult problem of determining the helium abundance in main-sequence 
stars \citep[e.g.][]{vau08,sv10}. The luminosity of stellar models on the 
subgiant branch is mainly determined by the amount of energy produced at 
the edge of the helium core established during the main-sequence phase. 
The rate of energy production depends on the temperature and the hydrogen 
abundance in that layer. As a consequence it is possible to find models at 
the same luminosity with quite different values of total mass, by 
adjusting the other parameters to yield the required temperature at the 
edge of the helium core. Fortunately, for subgiant stars the presence of 
mixed modes can provide additional constraints on the size of the helium 
core. For KIC~11026764 this mixed mode constraint significantly reduces 
the range of possible masses to two specific intervals, where the 
combination of stellar mass and core size are compatible with the 
atmospheric parameters and the frequencies of the mixed modes.

Regardless of which family of solutions is a better representation of 
KIC~11026764, the parameter values in Table~\ref{tab4} already yield 
precise determinations of some of the most interesting stellar 
properties---in particular, the asteroseismic age and radius. If we 
consider only the four models that were produced by the same modeling team 
with an identical fitting method (the ASTEC models: AA, AA$^\prime$, AB, 
AB$^\prime$), we can calculate the mean value and an internal 
(statistical) uncertainty in isolation from external (systematic) errors 
arising from differences between the various codes and methods. The 
asteroseismic ages of the two families of solutions range from 5.87 to 
5.99~Gyr, with a mean value of $t=5.94\pm0.05$~Gyr. The stellar radii of 
the two families range from 2.03 to 2.08~$R_\odot$, with a mean value of 
$R=2.05\pm0.03\ R_\odot$. The luminosity ranges from 3.45 to 
3.80~$L_\odot$, with a mean value of $L=3.56\pm0.14\ L_\odot$. These are 
unprecedented levels of precision for an isolated star, despite the fact 
that the stellar mass is still ambiguous at the 10\% level. Of course, 
{\it precision} does not necessarily translate into {\it accuracy}, but we 
can evaluate the possible systematic errors on these determinations by 
looking at the distribution of parameter values for the entire sample of 
modeling results, not just those from ASTEC.

Considering the full range of the best models ($\chi^2<10$) in 
Table~\ref{tab4}, there is broad agreement on the value of the stellar 
radius. The low value of 2.03~$R_\odot$ is from one of the ASTEC models 
considered above, while the high value of 2.09~$R_\odot$ is from the 
Catania-GARSTEC code, leading to a systematic offset of $^{+0.04}_{-0.02}\ 
R_\odot$ compared to the mean ASTEC value. There is a slightly higher 
dispersion in the values of the asteroseismic age. Again considering only 
the best models, we find a full range for the age as low as 4.99~Gyr from 
Catania-GARSTEC up to 5.99~Gyr from ASTEC, for a systematic offset of 
$^{+0.05}_{-0.95}$~Gyr relative to the ASTEC models. The best models 
exhibit the highest dispersion in the values of the luminosity, with a low 
estimate of 3.45~$L_\odot$ from ASTEC and a high value of 4.44~$L_\odot$ 
from Catania-GARSTEC. These models establish a systematic offset of 
$^{+0.88}_{-0.11}\ L_\odot$ compared to the ASTEC results.

To evaluate the relative contribution of each observational constraint to 
the final parameter determinations, we can study the significance $S_i$ 
using singular value decomposition \citep[SVD; see][]{bro94}. An 
observable with a low value of $S_i$ has little influence on the solution, 
while a high value of $S_i$ indicates an observable with greater impact. 
The significance of each observable in determining the parameters of 
KIC~11026764 for Model AA is shown in Figure~\ref{fig4}. From left to 
right we show the spectroscopic constraints followed by the $l=0$, 1 and 2 
frequencies labeled with the reference radial order from Table~\ref{tab1}, 
and each group of constraints is shown in a different shade. It is 
immediately clear that the $l=1$ and $l=2$ frequencies have more weight in 
determining the parameters. If we examine how the significance of each 
spectroscopic constraint changes when we combine them with modes of a 
given spherical degree, we can quantify the impact of each set of 
frequencies because the information content of the spectroscopic data does 
not change. The spectroscopic constraints contribute more than 25\% of the 
total significance when combined with the $l=0$ modes (13\% from the 
effective temperature alone), indicating that these two sets of 
constraints contain redundant information. By contrast, the total 
significance of the spectroscopic constraints drops to 4\% for the $l=2$ 
modes, and 7\% when combined with the $l=1$ modes---confirming that these 
frequencies contain more independent information than the $l=0$ modes, as 
expected from the short evolutionary timescale for mixed modes. Although 
the significance of the atmospheric parameters on the final solution 
appears to be small, we emphasize that accurate spectroscopic constraints 
are essential for narrowing down the initial parameter space.\\

%%%%%%%%%%%%%%%%%%%%%%%%%%%%%%%%%%%%%%%%%%%%%%%%%%%%%%%%%%%%%%%%%%%%%%%%%%%

\section{SUMMARY \& DISCUSSION\label{SEC6}}

We have determined a precise asteroseismic age and radius for 
KIC~11026764. Although no planets have yet been detected around this star, 
similar techniques can be applied to exoplanet host stars to convert the 
relative planetary radius determined from transit photometry into an 
accurate absolute radius---and the precise age measurements for field 
stars can provide important constraints on the evolution of exoplanetary 
systems. By matching stellar models to the individual oscillation 
frequencies, and in particular the $l=1$ mixed mode pattern, we determined 
an asteroseismic age and radius of 
$t=5.94\pm0.05$(stat)$^{+0.05}_{-0.95}$(sys)~Gyr and 
$R=2.05\pm0.03$(stat)$^{+0.04}_{-0.02}$(sys)~$R_\odot$. This represents an 
order of magnitude improvement in the age precision over pipeline 
results---which fit only the mean frequency separations---while achieving 
comparable or slightly better precision on the radius (cf. Models 
J$^\prime$ and K$^\prime$ in Table~\ref{tab4}). The systematic 
uncertainties on the radius are almost negligible, while the 
model-dependence of the asteroseismic age yields impressive accuracy 
compared to other age indicators for field stars \citep[see][]{sod10}. 
Whatever the limitations on absolute asteroseismic ages, studies utilizing 
a single stellar evolution code can precisely determine the {\it 
chronology} of stellar and planetary systems.

The $l=1$ mixed modes in KIC~11026764, shifted from regularity by avoided 
crossings, play a central role in constraining the models. Bedding et 
al.~(in prep.) have pointed out the utility of considering the frequencies 
of the avoided crossings themselves, since they reflect the g-mode 
component of the eigenfunction that is trapped in the core 
\citep{Aizenman77}. The avoided crossing frequencies are revealed by the 
distortions in the $l=1$ modes, which are visible as rising branches in 
Figure~\ref{fig1}. For Model AA, marked by the vertical line, we see that 
the frequencies of the first four avoided crossings are $G_1 \approx 
1270\,\mu$Hz, $G_2 \approx 920\,\mu$Hz, $G_3 \approx 710\,\mu$Hz and $G_4 
\approx 600\,\mu$Hz. Each of these avoided crossings produces a 
characteristic feature in the \'echelle diagram that can be matched to the 
observations. Indeed, the observed power spectrum of KIC~11026764 
(greyscale in Figure~\ref{fig3}) shows a clear feature that matches $G_2$ 
and another, slightly less clear, that matches $G_3$. The avoided crossing 
at $G_1$ that is predicted by the models lies outside the region of 
detected modes, but it is possible that additional data expected from the 
{\it Kepler Mission} will confirm its existence. We also note a peak in 
the observed power spectrum at 586~$\mu$Hz (greyscale in 
Figure~\ref{fig3}) that lies close to an $l=1$ mixed mode in the models, 
and also at 766~$\mu$Hz near an $l=2$ mixed mode. Again, additional data 
are needed for confirmation.

It is interesting to ask whether all of the models discussed in this paper 
have the same avoided crossing identification.  For example, are there any 
models that fit the observed frequencies but for which the avoided 
crossing at $920\,\mu$Hz corresponds to $G_1$ instead of $G_2$?  This 
would imply a different local minimum in parameter space, and a different 
location in the p-g diagram introduced by Bedding et al. Although this may 
be the case for some of the models in Table~\ref{tab3} from the initial 
search, all of the models listed in Table~\ref{tab4} have the avoided 
crossing identification described above.

The observed structure of the $l=1$ ridge suggests relatively strong 
coupling between the oscillation modes. At frequencies above the observed 
range, the best models suggest that the unperturbed $l=1$ ridge would 
align vertically near 40~$\mu$Hz in the \'echelle diagram (see 
Figure~\ref{fig3}). It is evident from Figure~\ref{fig1} that at a given 
age the frequencies of numerous p-modes are affected by the rising g-mode 
frequency, and this manifests itself in the \'echelle diagram with several 
modes deviating from the location of the unperturbed ridge on either side 
of the avoided crossing \citep{dm09}. Stronger coupling suggests a smaller 
evanescent zone between the g-mode cavity in the core and the p-mode 
cavity in the envelope. Although the evanescent zone will be larger for 
$l=2$ modes, the strength of the coupling for the $l=1$ modes raises the 
possibility that weakly mixed $l=2$ modes---like those near 766~$\mu$Hz in 
Model AA---may be observable in longer time series data from continued 
observations by {\it Kepler}.

It is encouraging that with so many oscillation frequencies observed, the 
impact of an incorrect mode identification appears to be minimal. For 
example, the models suggest that the lowest frequency $l=0$ mode in 
Table~\ref{tab1} may actually be on the $l=2$ ridge---or it could even be 
an $l=1$ mixed mode produced by the $G_4$ avoided crossing. However, the 
influence of the other observational constraints is sufficient to prevent 
any serious bias in the resulting models. Even so, adopting either of 
these alternate identifications for the lowest frequency $l=0$ mode would 
cut the $\chi^2$ of Model AB nearly in half.

Despite a 10\% ambiguity in the stellar mass, we have determined a 
luminosity for KIC~11026764 of 
$L=3.56\pm0.14$(stat)$^{+0.88}_{-0.11}$(sys)~$L_\odot$. With the radius so 
well determined from asteroseismology, differences of 200-300~K in the 
effective temperatures of the models are largely responsible for the 
uncertainties in the luminosity. These differences are generally 
correlated with the composition---hotter models at a given mass tend to 
have a higher helium mass fraction and lower metallicity, while cooler 
models tend to be relatively metal-rich. The adopted spectroscopic 
constraints fall in the middle of the range of temperatures and 
metallicities for the two families of models, and leave little room for 
substantial improvement. Perhaps the best chance for resolving the mass 
ambiguity, aside from additional asteroseismic constraints, is a direct 
measurement of the luminosity. Although {\it Kepler} was not optimized for 
astrometry, it will eventually provide high-quality parallaxes 
\citep{mon10}. The resulting luminosity error is expected to be dominated 
by uncertainties in the bolometric correction ($\sim$0.02 mag) and the 
amount of interstellar reddening ($\sim$0.01 mag), although saturation
from this bright target may present additional difficulties. This should 
lead to a luminosity precision near 3\%, which would be sufficient to 
distinguish between our two families of solutions for KIC~11026764.

%%%%%%%%%%%%%%%%%%%%%%%%%%%%%%%%%%%%%%%%%%%%%%%%%%%%%%%%%%%%%%%%%%%%%%%%%%%

\acknowledgments Funding for the {\it Kepler Mission} is provided by NASA's 
Science Mission Directorate. This work was supported in part by NASA grant 
NNX09AE59G. Computer time was provided by TeraGrid allocation TG-AST090107. 
The National Center for Atmospheric Research is sponsored by the U.S.~National 
Science Foundation. Observations were made with the Nordic Optical 
Telescope, operated on the island of La Palma jointly by Denmark, Finland, 
Iceland, Norway, and Sweden, in the Spanish Observatorio del Roque de los 
Muchachos of the Instituto de Astrofisica de Canarias. TA gratefully 
acknowledges support from the Programme National de Physique Stellaire of 
INSU/CNRS. GD, P-OQ, CK, JC-D and HK are grateful for financial support 
from the Danish Natural Science Research Council. J.M-\.Z acknowledges 
MNiSW grant N203 014 31/2650.  MJPFGM acknowledges financial support from 
project TDC/CTE-AST/098754/2008 from FCT \& FEDER, Portugal. TRB and DS 
acknowledge financial support from the Australian Research Council. WJC, 
YE, A-MB, STF, SH and RN acknowledge the support of the UK Science and 
Facilities Technology Council. OLC and P-OQ acknowledge support from 
HELAS, a major international collaboration funded by the European 
Commission's Sixth framework program. SGS acknowledges support from the 
FCT (Portugal) through grants SFRH/BPD/47611/2008 and 
PTDC/CTE-AST/66181/2006. The authors wish to thank the Kepler Science 
Team and everyone who helped make the {\it Kepler Mission} possible.

%%%%%%%%%%%%%%%%%%%%%%%%%%%%%%%%%%%%%%%%%%%%%%%%%%%%%%%%%%%%%%%%%%%%%%%%%%%

%%%%%%%%%%%%%%%%%%%%%%%%%%%%%%%%%%%%%%%%%%%%%%%%%%%%%%%%%%%%%%%%%%%%%%%%%%%
\appendix

  \section{A.\ THE INFLUENCE OF ROTATION\label{APPA}} %From Moya et al.
%MS \section{THE INFLUENCE OF ROTATION\label{APPA}} %From Moya et al.

Rotation velocity can have a large influence on the frequencies of a given 
stellar model, since angular momentum transport can change the internal 
structure and evolution of the star. Thus, rotation velocity and angular 
momentum transport processes must be taken into account for accurate 
modeling. However, if the rotation velocity is not too large, the effect 
on the frequencies can be comparable to or smaller than the observational 
accuracy, in which case non-rotating models can be trusted. We have 
examined two angular momentum transport processes covering the extreme 
cases to quantify the influence of rotation on the modeling of 
KIC~11026764.

The study was performed using the CESAM code \citep{Morel08}. The 
first-order effects of rotation on the equilibrium models were considered 
by subtracting the spherically averaged contribution of the centrifugal 
acceleration to the gravity of the model, $g_{\rm{eff}}=g-{\cal A}_c(r)$, 
where $g$ corresponds to the local gravity, and ${\cal A}_c(r)$ represents 
the radial component of the centrifugal acceleration. This spherically 
averaged component of the centrifugal acceleration does not change the 
order of the hydrostatic equilibrium equations. Such models are referred 
to as {\it pseudo-rotating} \citep[see][]{Soufi98, Sua07neadeg}. Since we 
have only a weak constraint on the rotation velocity of KIC~11026764 (see 
\S\ref{sec3.2}), we used an initial rotation velocity that leads to solar 
rotation at the solar age. Standard physical inputs were used, including 
the EFF equation of state. The opacity tables were taken from the OPAL 
package \citep{RogersIglesias95}, complemented at low temperatures 
($T\leq10^4\,\rm{K}$) with the tables provided by \cite{af94}. The outer 
boundary conditions were determined by assuming a plane-parallel Eddington 
grey atmosphere. The model metallicity $(Z/X)$ is derived from the 
observed [Fe/H] value assuming $(Z/X)_\odot=0.0245$ 
\citep{GrevesseNoels93}, $Y_{\mathrm{pr}}=0.235$ and $Z_{\mathrm{pr}}=0$ 
for the primordial helium and heavy-element abundances, and $\Delta 
Y/\Delta Z=2$ for the enrichment ratio. The thermonuclear reactions 
incorporated the PP and CNO cycles with the NACRE coefficients. No 
microscopic diffusion was included in the calculation.

We studied the following angular momentum transport processes: (1) Global 
Conservation of the angular momentum [GC, solid rigid rotation]: 
$\Omega(t,r)=\Omega(t)$, and (2) Local Conservation of the angular 
momentum [LC, differential rotation]: ${d\,r^2\Omega/dt}=0$. The GC and LC 
of angular momentum represent the two extreme cases in nature. The actual 
rotation profile of the star must fall between these two solutions.

The theoretical frequencies were calculated using GraCo \citep{graco2}. We 
found that for the rotation velocity studied here, rigid rotation yields 
frequencies closer to the non-rotating case, with differences in the range 
[$-$0.01, $-$0.025] $\mu$Hz. Differential rotation yields larger 
differences in the range [$-$0.051, $-$0.045] $\mu$Hz. Considering the 
large frequency separations, rigid rotation leads to differences relative 
to the non-rotating case in the range [$+$0.0025, $-$0.0045] $\mu$Hz while 
differential rotation yields differences in the range [0, $-$0.004] 
$\mu$Hz. All of these differences are much smaller than the observational 
errors, so rotation can safely be neglected in the frequency analysis of 
KIC~11026764. However, note that rotational mixing---which has not been 
considered in this study---may lead to changes in the global and internal 
properties of the models even for slowly rotating stars \citep{egg10}.\\

%%%%%%%%%%%%%%%%%%%%%%%%%%%%%%%%%%%%%%%%%%%%%%%%%%%%%%%%%%%%%%%%%%%%%%%%%%%

  \section{B.\ THE IMPORTANCE OF ASTEROSEISMIC CONSTRAINTS\label{APPB}} %From Quiron et al.
%MS \section{THE IMPORTANCE OF ASTEROSEISMIC CONSTRAINTS\label{APPB}} %From Quiron et al.

We compared results from SEEK with and without the asteroseismic inputs to 
see how the large and small separations can help reduce the uncertainty on 
the inferred stellar properties. We see that the uncertainty in the radius 
of 3\% provided by SEEK is in line with what was expected from simulated 
data \citep{Stello09a}. With $R = 2.10 \pm 0.06\ R_\odot$, the precision 
is a factor of five better than what we could get using only the available 
spectroscopic input ($T_{\rm eff}$, $\log g$ and [Fe/H]), which resulted 
in a 14.6\% uncertainty ($R=2.06 \pm 0.30\ R_\odot$). This dramatic 
improvement was obtained using the large and small separations as the only 
seismic constraints, as opposed to fitting individual mode frequencies.

% FIGURE 5 ---------------------------------------------------------------
  \begin{figure*}[t]
  \centerline{\includegraphics[angle=90,width=6.0in]{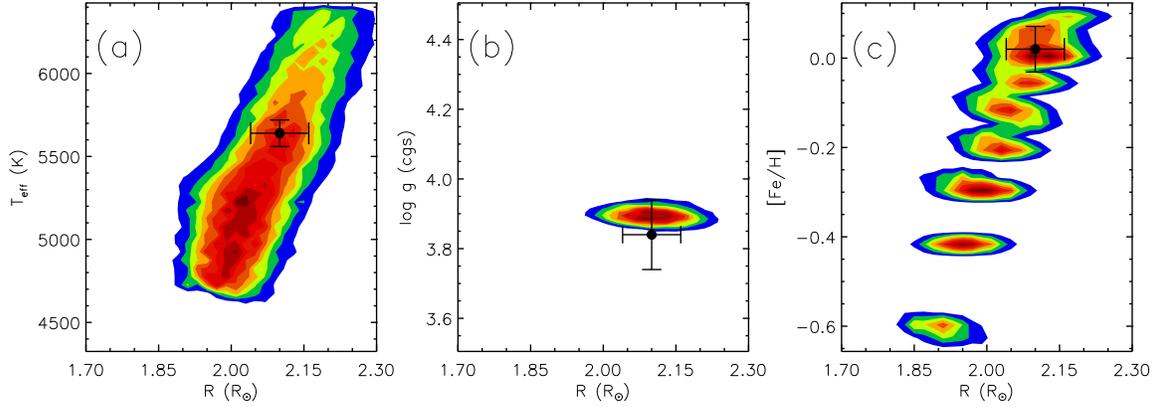}}
  \caption{Normalized probability distributions, showing the correlation of 
  the radius with several spectroscopic constraints when they are excluded 
  from the fit, including (a) the effective temperature $T_{\rm eff}$, (b) 
  the surface gravity $\log g$ and (c) the metallicity [Fe/H] relative to 
  solar, where the gaps between the islands are caused by the discrete 
  nature of the grid. Dark red indicates the maximum probability, which 
  decreases linearly to white for zero probability. In each panel, the cross 
  shows the observed values from the VWA method (see Table~\ref{tab2}) and 
  the final fit for $R=2.10\pm0.06$. We see that $\log g$ is determined 
  precisely before any spectroscopic knowledge of its value is included in 
  the fit.\label{fig5}}
  \end{figure*}
%-------------------------------------------------------------------------

In Figure \ref{fig5} we show 2D projections of the probability 
distributions provided by SEEK, which illustrate the correlations between 
the spectroscopic input parameters and the value of the inferred radius. 
To produce these figures we fixed all but one of the spectroscopic input 
parameters to the observed values. Hence, in Figure \ref{fig5}a the dark 
red regions are where we would most likely find the star if we did not 
know its temperature. The cross marks the observed value of the 
temperature with its measured uncertainty in the vertical direction while 
the horizontal direction marks the inferred value of $R$ and its 
precision. We can see that changing the temperature within the error bars 
does not greatly affect the value of the radius. Reducing the uncertainty 
of the temperature would not affect the uncertainty of the radius, since 
the width of the correlation function is large compared to its slope. In 
Figure \ref{fig5}b we see that the value of $\log g $ is well determined 
without any spectroscopic measurement. The spectroscopic value of $\log g 
$ has no significant influence in constraining the radius. The constraints 
from the seismic measurements are simply much stronger than what is found 
from spectroscopy in this case. We found similar results for the 
determination of the mass. Quite apparent in Figure \ref{fig5}c are 
several local maxima due to the discrete nature of the grid. However, the 
resolution of the grid is sufficient to see the underlying correlation 
between the metallicity and the radius. From this figure we see that if 
the star had a metallicity of [Fe/H]~$\sim -0.6$ it would have a radius of 
$R\sim1.91$~$R_\odot$. We remark that the uncertainty in radius is mainly 
caused by the unknown value of the initial helium content $Y_{\rm i}$ 
which has not been constrained by any of the observables \citep[for more 
details on the effect of $Y_{\rm i}$ on stellar parameters, 
see][]{QCDA10}.

% FIGURE 6 ---------------------------------------------------------------
  \begin{figure*}[b]
  \centerline{\includegraphics[angle=0,width=3.0in]{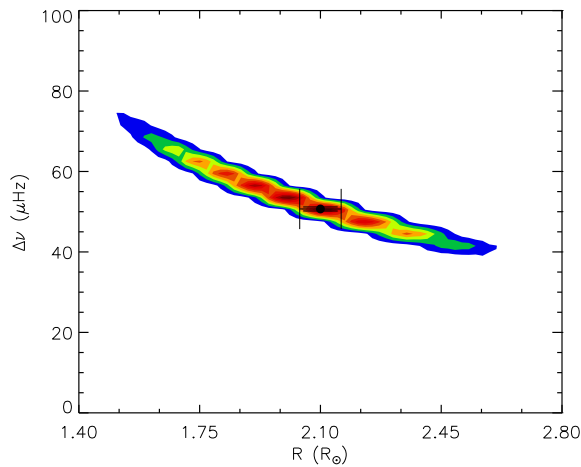}}
  \caption{Similar to Figure \ref{fig5}, but for $\Delta \nu_{0}$ when the 
  small separation is also excluded from the fit. The point with error bars 
  indicates the result when both the large and small separations are 
  included as constraints.\label{fig6}}
  \end{figure*}
%-------------------------------------------------------------------------

Finally, in Figure \ref{fig6} both the large and the small separations 
are left as free parameters. We note that the curved shape of the 
probability function is typical when the correlation between the large 
separation and the radius is plotted. In addition to showing that the 
large separation is the main constraint on the radius, Figure \ref{fig6} 
illustrates that a star with a large radius, and hence smaller value of 
the large separation, will have a larger uncertainty on the inferred 
radius.  The larger uncertainty is caused by the changing slope of the 
correlation, which tends to be flatter for larger radii. Figure 
\ref{fig6} also shows that without the asteroseismic constraints, the 
uncertainty in the radius increases dramatically since the most probable 
region for the radius, in red, spans $ 1.76 \la R/R_\odot \la 2.36 $.

For the mass, SEEK demonstrates that the precision is only slightly 
improved by including the large an small separations in the fitting 
process. We find $M= 1.25 \pm 0.13\ M_\odot$ when only the spectroscopic 
input parameters are used and $M=1.27\pm 0.09\ M_\odot$ when asteroseismic 
inputs are added. For the age, the results of SEEK without the 
asteroseismic inputs yield a relatively weak constraint of 40\% with 
$t=5.03\pm 2.02$~Gyr. This uncertainty is reduced by nearly a factor of 
two when the large and small separations are used, which 
gives $t = 4.26\pm 1.22$~Gyr. If we compare this result with 
\cite{QCDA10}, 30\% error on the age is unusually large for a star having 
well measured values of $\Delta \nu$ and $\delta \nu$. It seems that 
$\delta \nu$ does not provide an additional constraint on the age of the 
star, since we were able to determine a very similar age $t = 4.21\pm 
1.25$~Gyr when $\delta \nu$ was excluded from the fit. Bedding et al.~(in 
prep.) find this to be generally true for subgiant stars.

Finally, we stress that the mixing-length parameter $\alpha$ and the 
initial helium content $Y_{\rm i}$ are not constrained by our fit. For 
$Y_{\rm i}$, the output value is simply the central value used in SEEK's 
grid plus a 1$\sigma$ uncertainty extending toward the edges of the grid. 
For $\alpha$, a value smaller than 1.8 is preferred, but not decisively. 
Our overall conclusion from SEEK is that the asymptotic asteroseismic 
inputs can be used to reduce significantly the uncertainty on the radius 
of KIC~11026764. The age and the mass precision can also be improved, 
though less substantially.

\newpage

\end{document}